\documentclass[preprints,article,accept,pdftex,moreauthors, atoms]{mdpi}

\firstpage{1} 
\pubvolume{12}
\issuenum{12}
\articlenumber{71}
\pubyear{2024}
\copyrightyear{2024}
\externaleditor{Academic Editor: Chengjian Lin}
\datereceived{21 October 2024} 
\daterevised{19 November 2024} 
\dateaccepted{29 November 2024} 
\datepublished{19 December 2024} 
\hreflink{https://doi.org/10.3390/\linebreak atoms12120071} 

\usepackage[inkscapeformat=png]{svg}
\usepackage[format=plain]{caption}
\usepackage[labelformat=simple]{subcaption}
				
\DeclareCaptionLabelFormat{subcaptionlabel}{\normalfont(\textbf{#2}\normalfont)}				
\usepackage{braket}
\usepackage{float}
\usepackage{amsmath,mathtools}
\usepackage{graphicx}

\newcommand{\xenon}{$^{136}$Xe}

\newcommand{\bariumium}{$^{136}$Ba$^+$}

\newcommand{\msp}{\hspace*{-1.5mm}}

\Title{Ion Manipulation from Liquid Xe to Vacuum: Ba-Tagging \mbox{for a nEXO Upgrade and Future} $0\nu\beta\beta$ Experiments}

\TitleCitation{Ion Manipulation from Liquid Xe to Vacuum: Ba-Tagging for a nEXO Upgrade and Future $0\nu\beta\beta$ Experiments}

\Author{
{Dwaipayan Ray $^{1,}$*\orcidA{},} 
Robert Collister $^{2,}${*}, 
Hussain Rasiwala $^{3,}${*}\orcidB{}, 
Lucas Backes $^{1,4}$, 
Ali V. Balbuena $^{5}$, 
\mbox{Thomas Brunner $^{1,3}$\orcidC{},} 
Iroise Casandjian $^{1,4}$,
Chris Chambers $^{1,3}$\orcidD{}, 
Megan Cvitan $^{1,4}$, 
Tim Daniels $^{5}$, 
Jens Dilling {$^{6,\dagger,\ddagger}$,} 
Ryan Elmansali $^{2}$, 
William~Fairbank $^{7}$, 
Daniel Fudenberg $^{8,\mathsection}$, 
Razvan Gornea $^{2}$, 
Giorgio Gratta $^{8}$\orcidE{}, 
Alec Iverson $^{7}$, 
\mbox{Anna A. Kwiatkowski $^{1,9}$,} 
Kyle G. Leach $^{10,11}$\orcidF{}, 
Annika Lennarz $^{1,4}$, 
Zepeng Li $^{12}$, 
Melissa Medina-Peregrina $^{12}$\orcidG{}, 
\mbox{Kevin Murray $^{3,\parallel}$,}
Kevin O’Sullivan $^{8}$, 
Regan Ross $^{3}$,
Raad Shaikh $^{2}$\orcidH{}, 
Xiao Shang $^{3,\P}$,
Joseph Soderstrom $^{7}$, 
\mbox{Victor Varentsov $^{13}$\orcidI{}} 
 and Liang Yang $^{12}$\orcidJ{}
}

\AuthorNames{
Dwaipayan Ray, Robert Collister, Hussain Rasiwala, Lucas Backes, Ali V. Balbuena, Thomas Brunner, Iroise Casandjian, Chris Chambers, Megan Cvitan, Tim Daniels, Jens Dilling, Ryan Elmansali, William Fairbank, Daniel Fudenberg, Razvan Gornea, Giorgio Gratta, Alec Iverson, Anna A. Kwiatkowski, Kyle G. Leach, Annika Lennarz, Zepeng Li, Melissa Medina-Peregrina, Kevin Murray, Kevin O’Sullivan, Regan Ross, Raad Shaikh, Xiao Shang, Joseph Soderstrom, Victor Varentsov and Liang Yang
}

\AuthorCitation{{Ray, D.;}  Collister, R.; Rasiwala, H.; et al.}

\address{%
$^{1}$ \quad TRIUMF, Vancouver, BC V6T 2A3, Canada\\ 
$^{2}$ \quad Department of Physics, Carleton University, Ottawa, ON K1S 5B6, Canada\\
$^{3}$ \quad Physics Department, McGill University, Montreal, QC H3A 2T8, Canada\\
$^{4}$ \quad Department of Physics and Astronomy, McMaster University, Hamilton, ON L8S 4L8, Canada\\
$^{5}$ \quad Department of Physics and Physical Oceanography, University of North Carolina at Wilmington, \mbox{Wilmington, NC 28403, USA}\\
$^{6}$ \quad Department of Physics and Astronomy, University of British Columbia, Vancouver, BC V6T 1Z1, Canada\\
$^{7}$ \quad Physics Department, Colorado State University, Fort Collins, CO 80523, USA\\
$^{8}$ \quad Physics Department, Stanford University, Stanford, CA 94305, USA\\
$^{9}$ \quad Department of Physics and Astronomy, University of Victoria, Victoria, BC {V8P 5C2,} Canada\\
$^{10}$\quad Department of Physics, Colorado School of Mines, Golden, CO 80401, USA\\
$^{11}$\quad Facility for Rare Isotope Beams, Michigan State University, East Lansing, MI 48824, USA\\
$^{12}$\quad Department of Physics, University of California San Diego, La Jolla, CA 92093, USA\\
$^{13}$\quad Facility for Antiproton and Ion Research in Europe (FAIR GmbH), 64291 Darmstadt, Germany}

\corres{Correspondence: dray@triumf.ca (D.R.); rcollister@physics.carleton.ca (R.C.); \mbox{hussain.rasiwala@mail.mcgill.ca (H.R.)}} 

\firstnote{{Current address:} 
 Neutron Science Directorate, Oak Ridge National Laboratory, Oak Ridge, TN 37830, USA.}
\secondnote{{Current address:} 
 Physics Department, Duke University, Durham, NC 27708, USA.}
\thirdnote{Current address: Qventus Inc., Mountain View, CA 94043, USA.}
\fourthnote{Current address: Introspect Technology, Montreal, QC H3J 1M1, Canada.}
\fifthnote{Current address: Department of Materials Science and Engineering, University of Toronto, \mbox{Toronto, ON M5S 3E4, Canada}.}

\abstract{Neutrinoless double beta decay {($0\nu\beta\beta$)} provides a way to probe physics beyond the Standard Model of particle physics. 
The upcoming nEXO experiment will search for $0\nu\beta\beta$ decay in $^{136}$Xe with a projected half-life sensitivity exceeding $10^{28}$ years at the 90\% confidence level using a liquid xenon (LXe) Time Projection Chamber (TPC) filled with 5 tonnes of Xe enriched to $\sim$90\% in the {$\beta \beta$}-decaying isotope $^{136}$Xe. 
In parallel, a potential future upgrade to nEXO is being investigated with the aim to further suppress radioactive backgrounds and to confirm $\beta \beta$-decay events.
This technique, known as Ba-tagging, comprises extracting and identifying the $\beta \beta$-decay daughter $^{136}$Ba ion.
One tagging approach being pursued involves extracting a small volume of LXe in the vicinity 
of a potential $\beta \beta$-decay using a capillary tube and facilitating a liquid-to-gas phase transition by heating the capillary exit. The Ba ion is then separated from the accompanying Xe gas
using a radio-frequency (RF) carpet and RF funnel, conclusively identifying the ion as $^{136}$Ba
via laser-fluorescence spectroscopy and mass spectrometry.
Simultaneously, an accelerator-driven Ba ion source is being developed to validate and optimize this technique.
The motivation for the project, the development of the different aspects, along with the current status and results, are discussed here.
}


\begin{document}





\section{Introduction} \label{sec:intro}

The Standard Model (SM) of particle physics, which has been hugely successful in describing and even predicting subatomic particles and their interactions, was developed with the assumption that neutrinos are massless particles.
Observation of neutrino \mbox{oscillations~\cite{CAPOZZI:neutrino_oscillation2016218},} which necessitates a non-zero neutrino mass, points towards physics beyond the SM.
While their actual mass is still unknown, the limits set by experiments imply that neutrinos are at least six orders of magnitude smaller than electrons~\cite{aker2024directneutrinomassmeasurementbased}, which could suggest a different underlying mass-generation mechanism~\cite{Aker_2022_katrin_neutrino_nature, petcov2013adv_HEP}.
The fact that neutrinos are electrically neutral, massive particles opens the possibility that they are in fact Majorana particles, as proposed by E. Majorana~\cite{majorana_1937}. The most promising approach to probe for the Majorana nature of neutrinos is through searching for a unique decay process called neutrinoless double beta decay ($0\nu\beta\beta$)~\cite{.}. 
This $0\nu\beta\beta$ decay is expected to happen in addition to the SM-allowed two-neutrino double-beta decay ($2\nu\beta\beta$), which is a second-order weak nuclear process involving the simultaneous decay of two neutrons to two protons with the release of two electrons and two electron-anti-neutrinos. 
This rare type of decay has been observed in even-even nuclides where single $\beta$-decay is energetically forbidden~\cite{.}. 
If neutrinos are indeed Majorana particles, $0\nu\beta\beta$ will occur in these isotopes with the emission of only the two electrons, leading to a violation of lepton number.
This is required for leptogenesis~\cite{BUCHMULLER2005305} and subsequent explanation of the observed baryon asymmetry in the universe~\cite{Canetti_2012}.
In addition, $0\nu\beta\beta$ could also be able to shed light on the actual neutrino mass as well as elucidate the origin of such a small mass~\cite{majorana_1937,.}.

The EXO-200 experiment~\cite{Auger_2012_EXO200_Part1, Ackerman_2022_EXO200_Part2} has probed for $0 \nu \beta \beta$ decay in $^{136}$Xe using an active liquid xenon (LXe) mass of 110 kg, enriched to $80.6$\% in the $\beta \beta$-decaying isotope $^{136}$Xe at the Waste Isolation Pilot Plant (WIPP) underground facility in {Carlsbad,} 
 NM, USA, between 2011 and 2018. The experiment discovered $2 \nu \beta \beta$ decay in $^{136}$Xe~\cite{PhysRevLett.107.212501} and provided one of the most sensitive limits on the half-life of the $0\nu\beta\beta$ decay ($T_{1/2}>3.5 \times 10^{25}$\,yr at 90\% confidence level~(C.L.)~\cite{Anton:2019wmi}).
To increase this sensitivity, it is necessary to further suppress backgrounds (currently dominated by $\gamma$ rays) and increase the quantity of the parent isotope under observation. 
Based on the success of EXO-200, the next-generation experiment nEXO is under development, 
which will use a single-phase LXe Time Projection Chamber (TPC) approach with an active mass of 5 tonnes of Xe enriched to 90\% in the isotope $^{136}$Xe ~\cite{Alkharusi:arX}.
The projected sensitivity of nEXO to $0 \nu \beta \beta$ half-life in $^{136}$Xe is beyond $10^{28}$ years ($90\%$ C.L.) for 10 years of livetime, based on background estimates and signals from $2\nu\beta\beta$ decays~\cite{Adhikari_2022_sensitivity}.

\section{Barium Tagging for Future nEXO Upgrade}\label{sec:baTag}

While efforts continue to further reduce and suppress backgrounds in the nEXO experiment, parallel research is being conducted on new technologies and techniques, such as multivariate analysis or using electroformed copper, to increase detector \mbox{sensitivity~\cite{Adhikari_2022_sensitivity}.}
One such technique being developed for a potential future upgrade to nEXO is Ba-tagging~\cite{Moe_BaTagging_PhysRevC.44.R931}, that is, extracting and identifying the daughter isotope of barium, $^{136}$Ba, from the $\beta \beta$-decay of $^{136}$Xe. 
Since $^{136}$Ba is not produced in any of the $\gamma$-background events, Ba-tagging will eliminate all background events except those from $2\nu\beta\beta$ and will allow one to discriminate any event of interest as arising from a $\beta \beta$ decay or a background event.
In other words, a 100\% efficient tagging would allow a nEXO-type experiment to realize a sensitivity achievable in an almost background-free environment with just $2\nu\beta\beta$ decays, increasing the projected sensitivity by a factor of 2--3~\cite{Albert:2017hjq, Adhikari_2022_sensitivity} without increasing the isotope mass.

Due to the high reward from successful tagging, this technique is being pursued in different forms by several groups within the nEXO collaboration~\cite{RASIWALA2023298_emis, laser_spec, PhysRevA.91.022505, Chambers:2018srx, yvaine2024imaging} 
and outside~\cite{McDonald:2017izm}. Ba-tagging can be broadly divided into four main steps:
\begin{enumerate}
    \item Localisation: when a $\beta\beta$-like event in an energy window around the end-point energy ($Q_{\beta\beta} \approx 2.5$ MeV) is detected, the position of the decay within the detector is located. 
    \item \textls[-22]{Extraction: the Xe volume surrounding the event location is extracted from \mbox{the detector.}}
    \item Separation: a potentially present $^{136}$Ba ion is separated from the background Xe.
    \item Identification: the isolated $^{136}$Ba ion is positively identified.
\end{enumerate}

Step 1 has been demonstrated at EXO-200, where decay events were localized to within a few millimeters~\cite{albert_2014_PhysRevC.89.015502}. Step 4 has been achieved for different forms of Ba, for example, 
\mbox{Ba$^+$ trapped} in vacuum~\cite{laser_spec, laser_spec_thesis}, Ba atoms and ions trapped in a Xe-ice matrix~\cite{PhysRevA.91.022505, Chambers:2018srx, yvaine2024imaging}, and Ba$^{++}$ trapped in a molecule with Single Molecule Fluorescence Imaging (SMFI)~\cite{McDonald:2017izm}.
Depending on the technology of choice in Step 2 and Step 4, Step 3 becomes optional.
The current focus of Ba-tagging for nEXO is the demonstration of Step 2 (and 3 if necessary), and two main avenues are being investigated. 
Both comprise inserting a macroscopic external element, a cryoprobe in one and a capillary tube in the other approach, inside the LXe volume.
The cryoprobe approach, pursued by collaborators at Colorado State University{, Fort Collins, CO, USA} and the University of California{, San Diego, CA, USA,}
 entails trapping a Ba ion in a Xe ice matrix frozen to the tip of the probe, removing the probe from the detector volume with the Xe ice, and then using similar techniques to Ref.~\cite{PhysRevA.91.022505, Chambers:2018srx, yvaine2024imaging} for positive identification. 
The second extraction method involves flushing Ba ions out of the detector along with some LXe through a capillary, with subsequent transition to gas phase by heating the capillary exit.
This technique is being developed at Carleton University, Canada.
After extraction, the $^{136}$Ba ions are separated from the accompanying Xe gas using a radio-frequency (RF) carpet and RF funnel~\cite{brunner2015rf, brunner2017searching} and identified via laser fluorescence in a linear \mbox{Paul ion trap (LPT)~\cite{lan2020linear}} and mass spectrometry using a multi-reflection time-of-flight mass spectrometer (MRTOF-MS)~\cite{murray2023design}. These are being developed at McGill University, Canada.
Once a potential $0\nu\beta\beta$ event is detected, a positive Ba-tagging will unambiguously validate that event as a $\beta\beta$ event; otherwise, the event will be rejected as background.
This allows for a virtually background-free search for $0\nu\beta\beta$ without any {$\gamma$}
-photon contributions.
Thus, Ba-tagging, although challenging, holds great promise in providing an \mbox{irrefutable signal.}

\begin{figure}[H]
\vspace{1pt}

    \includegraphics[width=0.7\textwidth]{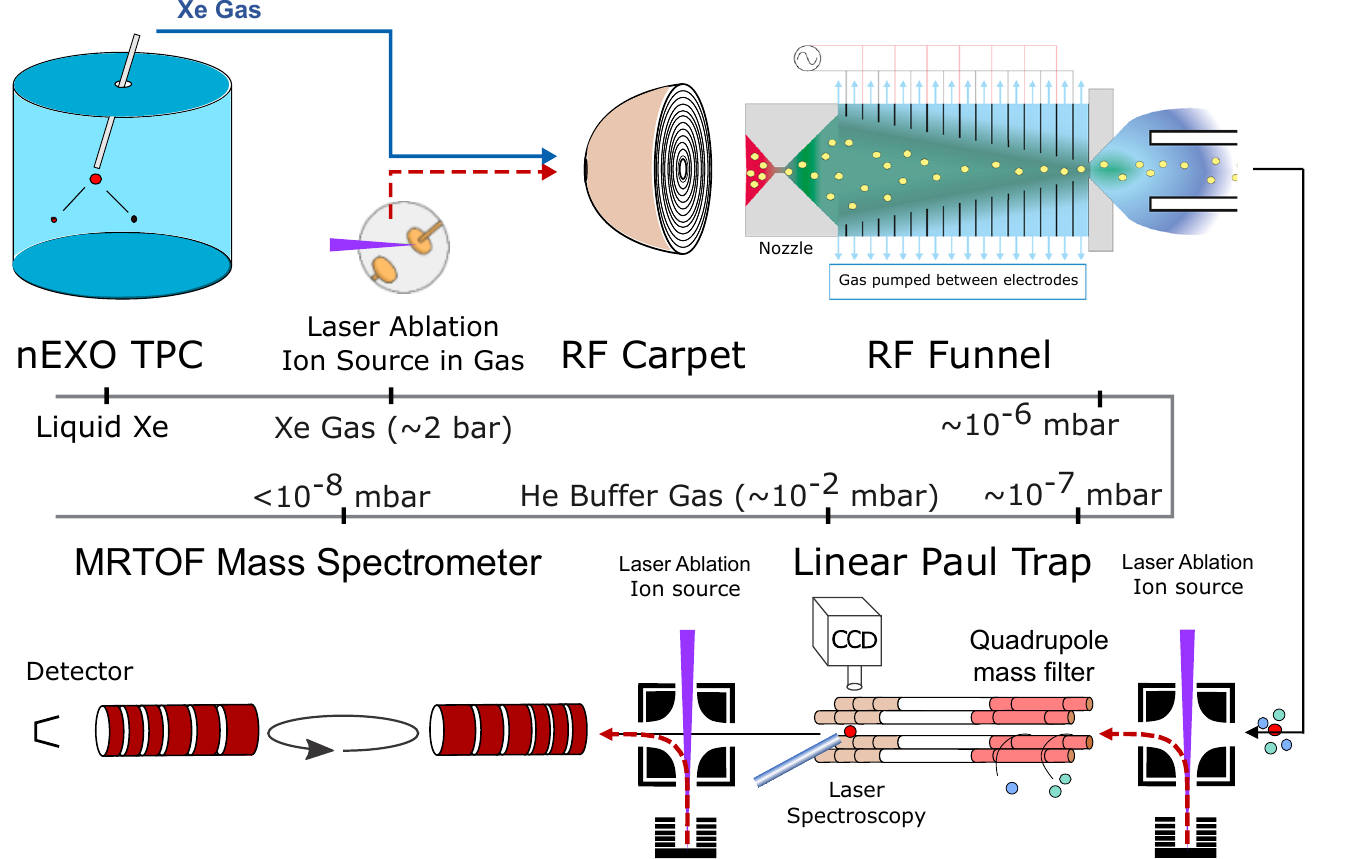}
    \caption{{Schematic} 
 depiction of the Ba-tagging approach involving extraction using a capillary tube, separation using an RF carpet and an RF funnel, and identification using laser-fluorescence spectroscopy in an LPT and MRTOF-MS. The respective pressures are shown in each step.}
    \label{fig:BaTag_stages}
\end{figure}

The Ba-tagging approach involving extraction using a capillary tube, separation using an RF carpet and an RF funnel, and identification using laser-fluorescence spectroscopy in an LPT and MRTOF-MS mass spectrometry is shown in Figure~\ref{fig:BaTag_stages} and is the focus of this paper. Section \ref{sec:indi_dev} describes the individual aspects of this approach, from the extraction of the Ba ions to their detection. 
To demonstrate the feasibility of the entire procedure of Ba-tagging, a Ba ion source is required. An accelerator-driven ion source is currently being developed at TRIUMF, Canada, where radioactive ions will be stopped in LXe, extracted electrostatically, and identified using $\gamma$~spectroscopy. The details of this development is discussed in Section \ref{sec:ion_source}. 
The ion source will lead the way for future ion source developments, which will be used to optimize either the cryoprobe or the capillary approaches before either is implemented in nEXO-type detectors in the future. 

\section{Progress on Ba-Tagging Subsystems}\label{sec:indi_dev}

Each step of the Ba-tagging approach shown in Figure~\ref{fig:BaTag_stages} will be discussed in detail in this section, highlighting recent progress in each area under development. The work is presented in order of the steps a Ba ion would experience following a decay event. 
First, the progress at Carleton University will be discussed, beginning with the LXe TPC for decay identification. This will include the light and charge collection systems and the capillary probe for extracting individual ions from the TPC, along with the displacement device and heating system to control the phase change in the capillary. 
Second, the developments at McGill University will cover the RF funnels for separating the ion from the xenon gas, followed by the ion identification stages for mass filtering and time-of-flight spectrometry.
Additionally, many of the tools developed to aid in these endeavors will be presented, such as the ion cooler and buncher that are necessary to demonstrate the full resolving power of the MRTOF-MS.

\subsection{Ion Collection and Extraction from LXe}
\label{sec:capillary}

In a LXe TPC, the collection of light and charge in coincidence can measure the total energy of a decay. It can also locate the site of the decay based on where the charge is collected and the time between the detected light and charge pulses. LXe is a high-yield scintillator with a peak emission at 175 nm. In the EXO-100 TPC at Carleton, the Xe scintillation light is wavelength-shifted by a tetraphenyl butadiene-coated Teflon reflector to a blue spectrum peaked at 430 nm before being collected by four photomultiplier tubes (PMTs). The collection of this prompt light provides a $t$ = 0 for the event. In modest electric fields of approximately 500 V/cm, the electrons generated by the ionization track travel at 1.71 mm/{$\upmu$}s \cite{albert_2014_PhysRevC.89.015502} through the 170 mm diameter, 120 mm long cylindrical drift volume. These electrons drift directly to the anode, where a pair of crossed wire planes (63 wires spaced 2 mm apart per plane) provide two coordinates for the location of the event. The third coordinate is given by the drift time and drift velocity. The total energy of the event, given by the combined magnitudes of the charge and light pulses, as well as the characteristics of the location information, e.g., the length of tracks and whether it is a single-site or multiple-site event, determines the nature of the decay as well as the species involved. This determines if a decay event is of interest for ion extraction and where to aim the probe to extract the daughter ion.

\begin{figure}[H]
		\includegraphics[width=0.49\textwidth]{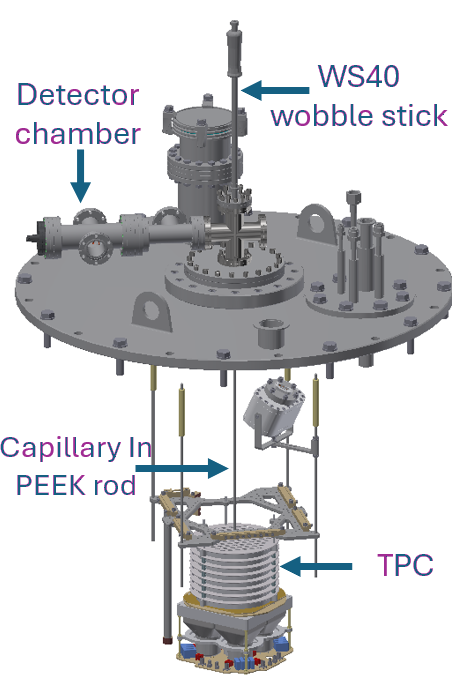}
		\includegraphics[width=0.49\textwidth]{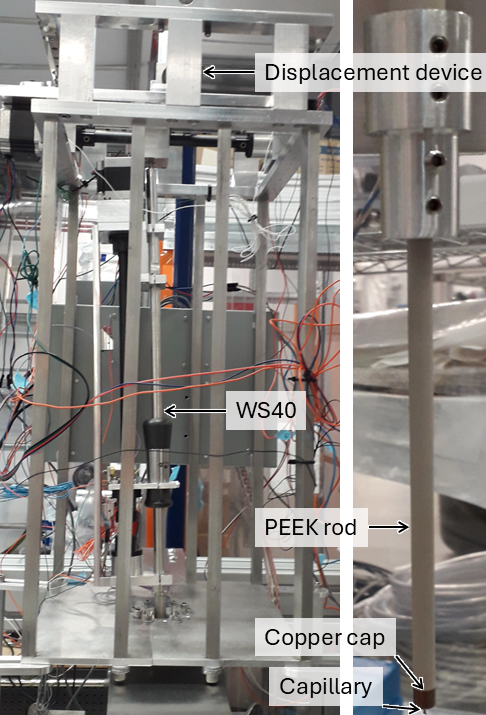}
	\caption{(\textbf{{Left}
}) CAD model of the TPC, probe, and detector chamber, without the displacement device. The stepper motors on the displacement device position the WS40 wobble stick. The internal shaft of the stick is magnetically coupled to its handle, enabling insertion and retraction of the capillary probe through the holes in the TPC cathode. (\textbf{{Right}
}) Photographs of the displacement device with WS40 in its test stand and the PEEK rod encapsulating the capillary. The copper cap on the PEEK rod is part of the homing system to calibrate the probe position.}
	\label{fig:cryostat}
\end{figure}

The heart of the probe is a commercial stainless steel capillary with a nominal inner diameter of 0.508 mm and a wall thickness of 0.2 mm. Laminar LXe flow up this capillary transports the ion out of the TPC into the subsequent stages of the tagging scheme. This capillary is encapsulated in a set of polyether ether ketone (PEEK) rods, providing the rigidity necessary to precisely position the capillary tip in a desired location. Flow along the capillary is controlled by manipulating the pressure differential across the inlet and outlet of the capillary, the inlet being in the cryostat, immersed in LXe with the TPC, and the outlet in a separate chamber either with a detector for initial development or the subsequent stages of the ion identification scheme. At the outlet, at a long distance from the LXe in the TPC, the capillary is heated to promote a phase change in the xenon from liquid to gas. The probe is mounted to a UHV Design brand WS40 wobble stick \cite{WS40}, which is manipulated by a custom-built displacement device consisting of three stepper motors on an aluminum frame, with two motors controlling the XY stage to set the probe angle and the third motor handling the insertion and retraction of the probe tip. The setup is shown in Figure \ref{fig:cryostat}. Once the device has been fed location information from online analysis of the TPC event signals, the displacement device is capable of positioning the probe tip anywhere within the accessible volume of the TPC within 15 s, with a precision of 0.11 mm on any axis. We also have a homing system consisting of six sets of pogo pins that allows calibration of the probe position at known locations above the TPC.
\vspace{-10pt}

COMSOL simulations were performed to study the xenon flow properties into, through, and out of the capillary. Laminar flow through the capillary is maintained as long as the average LXe extraction velocity is lower than 760 mm/s, corresponding to a Reynolds number below 2300. Above this, turbulence is introduced to the flow, which may direct a transiting ion into the capillary wall, where it could be neutralized and lost to the detection scheme. In laminar flow, the flow separates into layers, with a radial velocity profile that follows a smooth trend with a lower fluid velocity at the walls of the capillary and a larger fluid velocity in the center. This radial velocity profile is the critical boundary condition for the COMSOL simulations that were performed to study the ion extraction from LXe. These simulations solve the computational fluid dynamics of LXe being drawn up our capillary and determine the trajectories of a diffuse ion cloud located below the capillary tip in the LXe. From these simulations, it is found that only ions moving in the drift field directly into the capillary opening will be captured by the flow up the capillary; lateral displacements on the order of 0.5 mm prevent the ion from being collected. This is due to the distortion of the electric field by the conductive capillary; it redirects the field and drives the ions into the capillary walls. Alternative scenarios have been simulated, such as floating the capillary at a voltage or using a non-conductive PEEK capillary, but the drift field always dominates the motion except in the small volume with the strongest flow immediately around the capillary inlet. This greatly limits the capture volume, i.e., the volume of LXe where an ion must be located in order for it to be collected and carried up the capillary under these flow conditions. Hence the drift field must be turned off during ion collection for an acceptable extraction efficiency. 

\begin{figure}[H]
\vspace{2pt}
		\includegraphics[width=.98\textwidth]{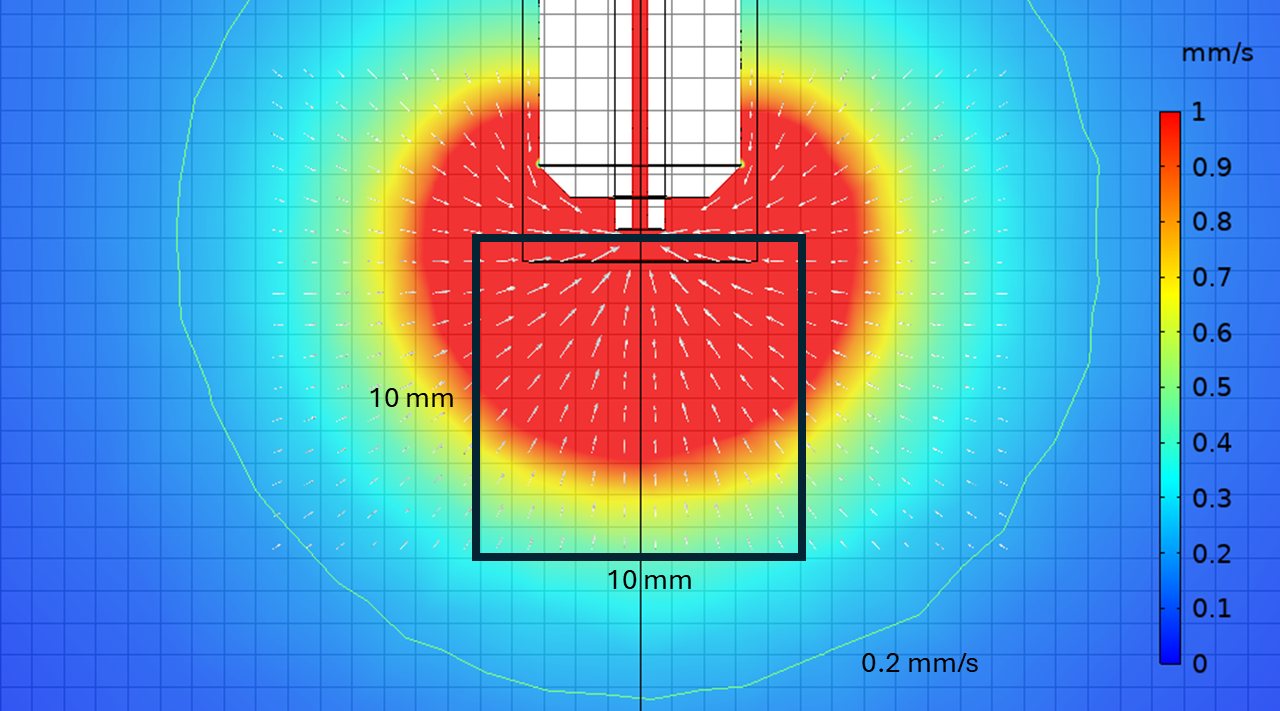}
	\caption{Simulation showing the flow map of the 0.508 mm inner diameter capillary encapsulated in the PEEK rod, with an inlet velocity of 750 mm/s and the electric drift field turned off, showing a slice of the flow field. The 10 mm square indicates the extent of the diffuse ion cloud generated for extraction studies. The contour indicates the 0.2 mm/s flow boundary at the edge of the capture~volume.}
	\label{fig:COMSOL_extraction}
\end{figure}

In this no-drift-field scenario, the size of the capture volume is determined by the flow rate against the random thermal drift of the ion. We conservatively estimate this random drift to be less than 0.2 mm/s based on Rn-Po coincidence measurements performed in the EXO-200 TPC~\cite{albert2015}; any ion inside the 0.2 mm/s flow region should eventually make it to the capillary tip. The measured ion mobility from EXO-200 of 0.219 cm$^2$/kVs may be used with the Einstein–Smoluchowski equation to calculate a diffusion constant and, assuming isotropic diffusion, a random drift velocity. This gives a drift velocity roughly 20 times smaller than the empirical value above. Since we are using radioactive ions in proof-of-concept measurements, capturing from a larger volume with slower flow near its extremities gives the ion more time to decay and hence be lost to our detection scheme before extraction. Thus we are content to use this empirical drift velocity to define our volume of interest. 
Figure~\ref{fig:COMSOL_extraction} shows a slice of the capture volume for the typical operation of the ion extraction probe, with an average extraction velocity of 750 mm/s to maintain laminar flow. This corresponds to an LXe flow of roughly 0.5 g/s, a negligible fraction of the total xenon volume of nEXO and easily replaced during recirculation. Also indicated is the 1~cm$^3$ volume where the diffuse cloud of regularly spaced ions is generated in simulation for extraction, investigating the proximity of the capillary to the event of interest required for efficient extraction. The outermost flow contour indicates where the flow has fallen to 0.2 mm/s, showing the outer edge of the capture volume. All of the ions within the 1 cm$^3$ cloud are extracted by the capillary within 30 s of its arrival at the collection position, and they are concentrated radially within the center 3/4 of the capillary. This indicates that if the probe tip is positioned within a few mm of the ion, the ion will be extracted regardless of any thermal drift. Thermal currents in the LXe are not included in this simulation, but this could in principle be accounted for by a flow map and appropriately targeting \mbox{the probe tip.}

Simulations of the moving capillary probe, consisting of the capillary and enclosing PEEK rod as they are moved at 10 mm/s into the LXe and brought to rest at a target depth, show that the agitation of the LXe is minimal in the target 1 cm$^3$. The largest introduced flow is immediately below the probe tip before it stops at the target position. At 1 mm below the probe tip, at the edge of the target 1 cm$^3$, the motion is on the order of 1.5 mm/s. At that position, the extracting flow is over 100 mm/s in the opposite direction. Elsewhere, and on the axes orthogonal to the probe's motion, the agitation is at least an order of magnitude smaller than the extraction flow, always in the opposite direction. Thus, the agitation introduced by the moving probe negligibly disrupts the target ion \mbox{during extraction.}

\subsection{Ion Transport Through a Capillary}

The next stage is the ion's transport through the capillary, where the LXe is heated to undergo a phase transition to gas. By using a metal capillary, we are able to apply a current through its body to deliver heat in a well-defined location, with regulation of the transferred heat by Proportional Integral Derivative (PID) control. The apparatus includes a pair of copper clamps, necessary to make a robust electrical connection to the thin capillary, and resistance temperature detectors (RTDs) epoxied to each clamp with their own connections. The temperature difference between the clamps is monitored to ensure the phase transition is complete. This heated section of capillary is roughly 80 cm from the LXe in the TPC and on the other side of a nominally room temperature flange. LXe flowing up the capillary over this distance prevents any heat transfer backwards to the LXe in the TPC.

\begin{figure}[H]
\vspace{2pt}
		\includegraphics[width=0.33\textwidth]{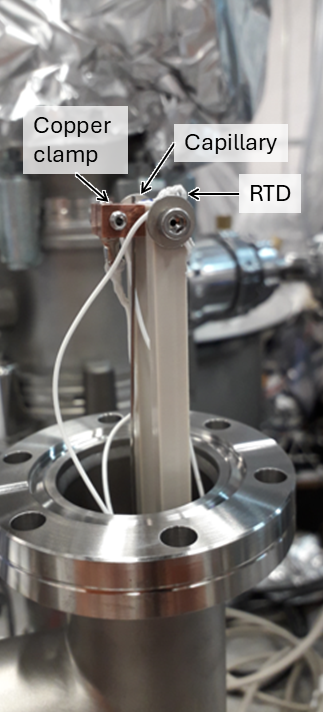}
		\includegraphics[width=0.66\textwidth]{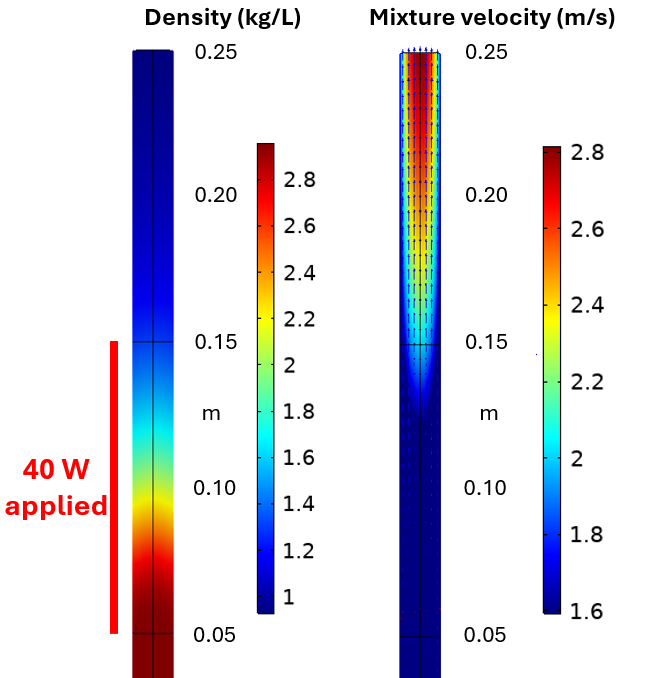}
	\caption{(\textbf{{Left}
}) Image of the capillary showing one of the heating clamps with RTD. \mbox{(\textbf{{Middle}
})~COMSOL} simulation result of the density, showing where the heat is applied and the resulting density gradient as the phase transition occurs. (\textbf{{Right}
}) COMSOL simulation result of the mixture velocity, showing the last length of capillary, demonstrating that the flow does not significantly deviate from laminar as the phase transition occurs.}
	\label{fig:phase_change}
\end{figure}

COMSOL computational fluid dynamics and heat transfer simulations inform if an ion survives the journey up the capillary, including the phase transition, without touching the wall. We employ a mixture model where the xenon is treated as a dispersed phase (gas) in a continuous phase (liquid). For the heat transfer, the mixture is treated as a single fluid with properties given by a linear combination of the gas and liquid fractions' properties. For example, for the density $\rho$,
\begin{equation}
\mu = X_\textrm{g} \rho_\textrm{g} + X_\textrm{l} \rho_\textrm{l},
\end{equation}
where $X_{\alpha}$ is the phase fraction and $\rho_{\alpha}$ the phase's density, with the subscript $\alpha = \textrm{g,l}$ denoting the gas or liquid phase, respectively. Similarly, the thermal conductivity and heat capacity are defined. Xenon transitions between phases by the balance of the evaporation and condensation rates, which are influenced by how much heat is added to the system. For the setup discussed here, 40 W are applied over a 0.1 m length immediately before the outlet of the capillary. However, in the simulation, an additional 0.1 m of capillary are added after the heating region to observe how the mixture will continue to evolve. \mbox{Only 0.05 m} before the heat is applied is simulated; before this, in the leading section of the 1 m long capillary, there is only uniform laminar flow. Particle trajectories are traced through the resulting flow map along the length of the capillary. Some results are shown \mbox{in Figure~\ref{fig:phase_change}.}

 The temperature of the mixture at the exit of the capillary is roughly 190 K, well above the boiling point of LXe, suggesting that enough energy for a complete phase change was provided. Once the mixture emerges into the relatively lower pressure of a detector chamber, or the RF funnel of the tagging scheme, any remaining superheated liquid droplets should finish transitioning to gas. The particle trajectories produced by the simulation show some ion loss from a small amount of turbulence introduced within about \mbox{0.02--0.03 mm} of the capillary walls, indicative of boiling. Fortunately, the extraction simulations previously discussed suggest that these layers of the flow will not be populated by ions if the probe is accurately targeted to a decay event in the TPC. Thus, this simulation suggests that ion transfer should be close to 100\% efficient for our scheme once the ion is inside \mbox{the capillary.}

\subsection{Ion Extraction from Gaseous Xenon}
\label{sec:funnel}

\textls[-15]{In liquid xenon, it is expected that the $^{136}$Ba$^{++}$ would gain an electron from a xenon atom to become $^{136}$Ba$^{+}$ due to the different ionization potentials~\cite{Moe_BaTagging_PhysRevC.44.R931}. Extraction of this \bariumium\space using the capillary tube will result in the ion being in a \mbox{high-pressure gaseous Xe (GXe)}} environment. In this condition, the motion of ions is dominated by collisions with the GXe. To facilitate the extraction and identification of \bariumium\space \msp, the Ba-tagging scheme uses an RF-only ion funnel to guide and separate the ions from the GXe~\cite{brunner2015rf, RASIWALA2023298_emis}. RF funnels, or ion funnels in general, are devices used to focus beams of ions; first devised for use with electrospray ionization mass spectrometry (ESI-MS)~\cite{page2008subambient} at a pressure of 30 torr. Since then, ion funnels have been used in several mass spectroscopy applications involving ionization sources as well as gas-filled stopper cells at fragmentation facilities~\cite{wada2013, varentsov2023review}.

\begin{figure}[H]
    \includegraphics[width=0.99\textwidth]{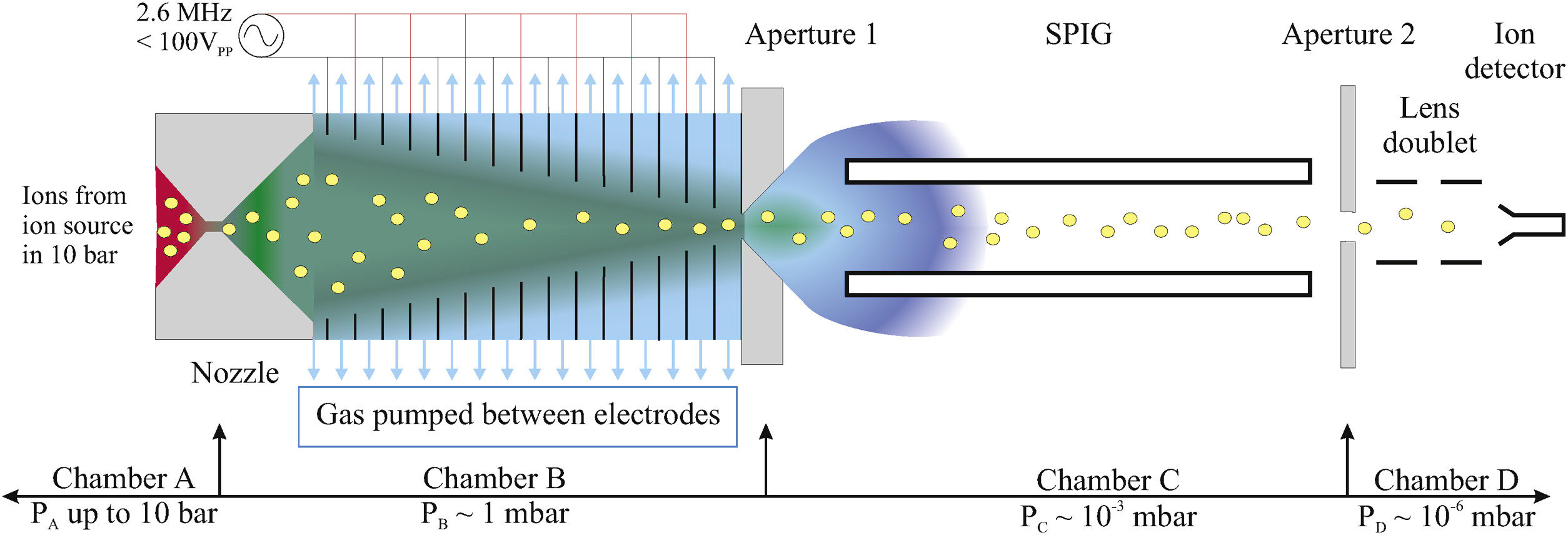}
    \caption{{Schematic}
 diagram of the RF funnel setup. Ions are injected through the converging–diverging nozzle at pressures of up to 10 bar. Injected ions are guided by the RF potentials applied to the annular disc electrodes of the RF funnel, while GXe is pumped out from between the electrodes. Ions then propagate through the SPIG and past Aperture 2 to be finally detected. Different ion detection setups were used downstream of Aperture 2. Taken from~\cite{brunner2015rf}.}
    \label{fig:funnel_old_schematic}
\end{figure}

The RF-only ion funnel for Ba-tagging, as shown in Figure~\ref{fig:funnel_old_schematic}, is designed to accept \bariumium\space from the capillary in GXe pressures of up to 10 bar and transport it to a region of high vacuum~\cite{brunner2013setup}. Stacked annular disc electrodes with tapering hole sizes, supplied with RF potential, radially confine ions by restricting motion away from the axis. Due to the RF-only design, there are no axial drift fields, and the ion propagation along the funnel axis is carried out primarily through the residual gas flow near the axis. This reduces the possibility of contaminants in the vacuum system by eliminating the need for components such as resistor chains for a DC drag field.
\vspace{-4pt}

\begin{figure}[H]
    \includegraphics[width=0.99\textwidth]{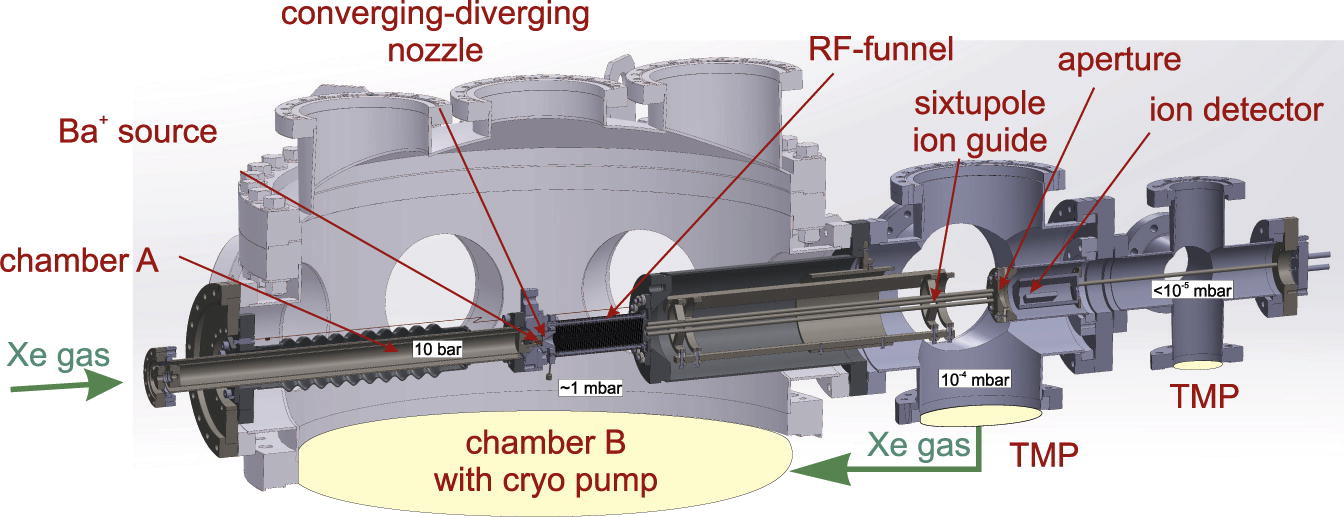}
    \caption{{CAD} 
 model rendering of the RF funnel test setup. A $^{148}\text{Gd}$-driven Ba-ion source generates ions for supporting the commissioning. The RF funnel and other ion optics are located inside a cryopump chamber for Xe gas recovery, and turbomolecular pumps (TMP) evacuate \mbox{chambers C and D} \mbox{in Figure~\ref{fig:funnel_old_schematic}.} Taken from~\cite{brunner2013setup}.}
    \label{fig:funnel_old_pic}
\end{figure}

The design and initial results of the RF funnel are reported in~\cite{brunner2015rf}. A $^{148}\text{Gd}$-driven Ba-ion source was used to generate $\text{Ba}^+$ ions~\cite{montero2010simple} upstream from a converging–diverging nozzle. The nozzle injects the ions with carrier gas (GXe) into the RF funnel, which is located inside a cryopump chamber for recovery of the GXe. Ions from the RF funnel travel through a sextupole ion guide (SPIG) to a channel electron multiplier (CEM) where they are detected. The SPIG stage is required to reduce the pressure after the RF funnel stage further to achieve a $10^{-6}$ mbar pressure that is favorable for operating the CEM \mbox{(see Figure~\ref{fig:funnel_old_schematic}).} The GXe captured in the cryopump during this operation is recovered by and stored in a gas-handling system for future use. Figure~\ref{fig:funnel_old_pic} shows the initial RF funnel setup at \mbox{Stanford University.}

\begin{figure}[H]
    \includegraphics[width=0.99\textwidth]{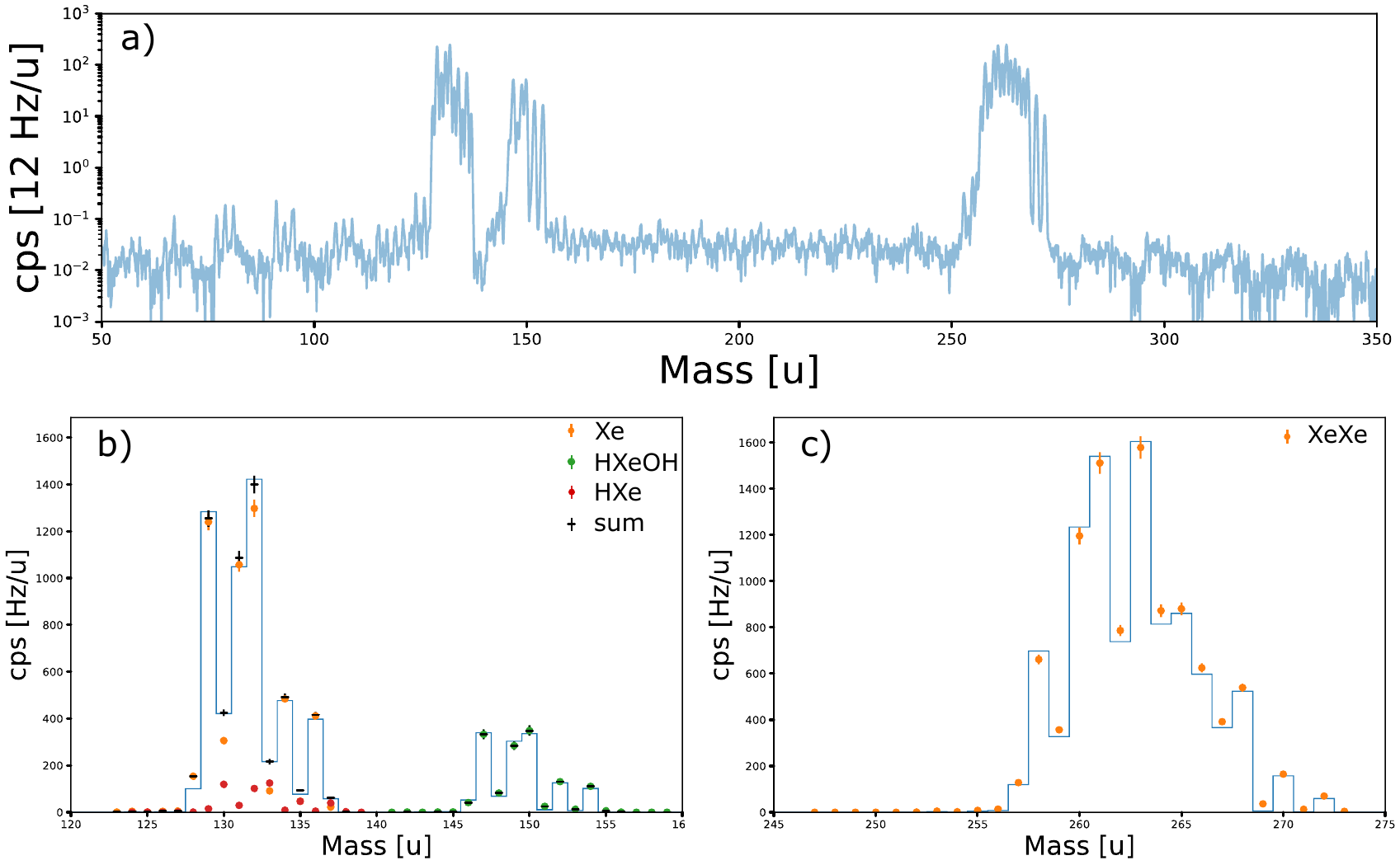}
    \caption{{(\textbf{a})} 
 The m/q spectrum of ionized entities extracted from a $^{252}\text{Cf}$ source at 2.1 bar GXe pressure. (\textbf{b},\textbf{c}) Fits of this spectrum to possible singly charged molecular entities. Aside from no evidence of fission products, the spectrum shows the presence of singly charged Xe and molecules such as xenon dimers (XeXe), xenon hydride-hydroxide (HXeOH), and possibly xenon hydride (HXe). Taken from~\cite{fudenberg2018improved}.}
    \label{fig:dan_plot}
\end{figure}

Ion transmission efficiency as a function of RF potential at different pressures showed similarities in trends at high RF potentials across pressures from 2--10 bar when compared with simulations~\cite{brunner2015rf}. Although this setup was successful in validating the simulated transmission efficiencies, it lacked ion discrimination capability. This proved to be important since the $^{148}\text{Gd}$-driven Ba-ion source also generated $\alpha$-particles that ionized the Xe atoms. Thus, to allow for the identification of the extracted ions, a {commercial}
\endnote{Thermo Finnigan LTQ-FT ICR, {Thermo Fisher Scientific, Waltham, MA, USA}.} 
 linear \mbox{quadrupole ion trap (LTQ)} was introduced~\cite{schwartz2002two, fudenberg2018improved}. Additionally, a $^{252}\text{Cf}$ source was used in place of the $^{148}\text{Gd}$ source to achieve a higher ion flux for better calibration of the LTQ. 
Figure~\ref{fig:dan_plot} shows ions extracted from the $^{252}$Cf source placed upstream of the nozzle in 2.1 bar of Xe pressure. The spectrum is normalized to counts per second, binned to integer values, and fit to the isotopic signatures of various molecular species. The natural abundance of elements in these molecules is used to calculate their m/q value. The bin corresponding to this value is 5\% smeared into the adjacent bins to account for the spectrometer's resolution. Fitting molecule signatures to the observed spectrum showed clear evidence of the presence of singly charged Xe ions and molecules of xenon, such as XeXe (dimer), HXeOH (xenon hydride-hydroxide), HXe (xenon hydride), and less than 100 ions ($\approx$1$\%$ of the total number of ions detected) that could be $^{252}\text{Cf}$ fragmentation products.    

\vspace{-15pt}

\begin{figure}[H]
    \includegraphics[width=\textwidth]{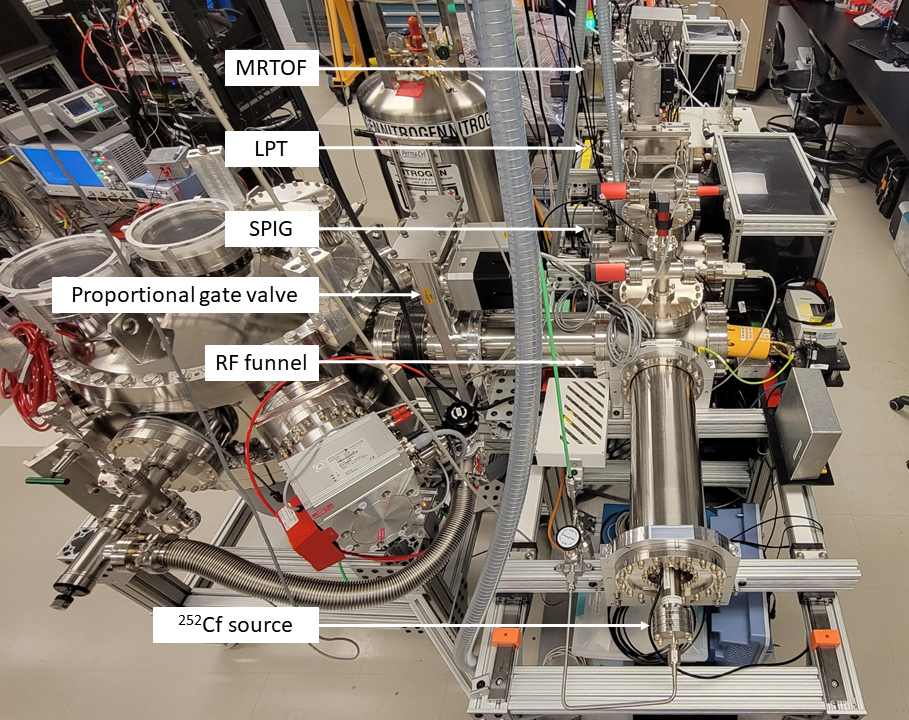}
    \caption{Picture of the Ba-tagging apparatuses being commissioned at McGill University for studying ion extraction from GXe. Vacuum system shown here houses the RF funnel, which connects to the cryopump chamber via a proportional gate valve. Located downstream from the RF funnel are the LPT and the MRTOF-MS.}
    \label{fig:ibis}
\end{figure}

The RF funnel setup is currently being recommissioned at McGill University, with the RF funnel now located in a dedicated vacuum chamber that is connected to the cryopump for Xe gas recovery. This was done to prevent thermal contractions of the RF funnel that were suspected as the cause for the observed capacitance changes when installed on top of the cryopump. A photograph of the setup is shown in Figure~\ref{fig:ibis}. The cryopump chamber couples to the RF funnel through a proportional gate valve for better control over the RF funnel backing pressure. 
The proportional gate valve will be operated with feedback from pressure gauges installed on the RF funnel chamber for precise pressure control. This feedback control is currently being developed into a LabVIEW program and will be first tested with Ar gas and later with GXe. 
The Ba-tagging scheme is further extended to replace the LTQ with a custom-designed LPT~\cite{lan2020linear} and an MRTOF-MS~\cite{murray2023design}. The LPT traps ions to allow for Ba ion identification via laser fluorescence spectroscopy~\cite{laser_spec}, while the MRTOF-MS will be used to achieve high mass resolving power (m$/\hspace{-0.5mm}\Delta$m $\approx$ 100,000) to discriminate $^{136}\text{Ba}^+$ from $^{136}\text{Xe}^+$ (compared with m$/\hspace{-0.5mm}\Delta$m $\approx$ 800 from the LTQ) and perform other systematic studies. Additionally, a quadrupole mass filter (QMF) is located upstream from the LPT to filter any dominant background ions from the RF funnel to improve \mbox{ion trapping.} 


As an upgrade to the apparatuses shown in Figure~\ref{fig:ibis}, a design is being proposed for developing an RF carpet, to be located upstream from the RF funnel. RF carpets are a popular choice for use in gas stopper cells~\cite{wada2003}, which are used in accelerator facilities to thermalize and collect ions. An application, similar to the one presented here, is being explored by the NEXT collaboration for Ba$^{++}$ detection using the SMFI technique \cite{jones2022dynamics}. 

The RF carpet will facilitate the transfer of $^{136}$Ba$^{+}$ coming from the capillary to the RF funnel. The incoming Ba ion (in GXe) drifts towards the RF carpet, which then guides the ion towards the exit aperture.   
For ion propagation along the carpet, an ion-surfing mode is being investigated that utilizes a combination of an RF potential and a high-frequency potential to repel and guide the ions, respectively, as described in~\cite{bollen2011ion}. Once developed, the RF carpet will support the efficient transfer of \bariumium\space from the capillary exit to the RF funnel as shown in Figure~\ref{fig:BaTag_stages}.

\subsection{Ion Identification}
\label{sec:lpt_mrtof}
 
\begin{figure}[H]
   \hspace{-4pt} \includegraphics[width=0.96\textwidth]{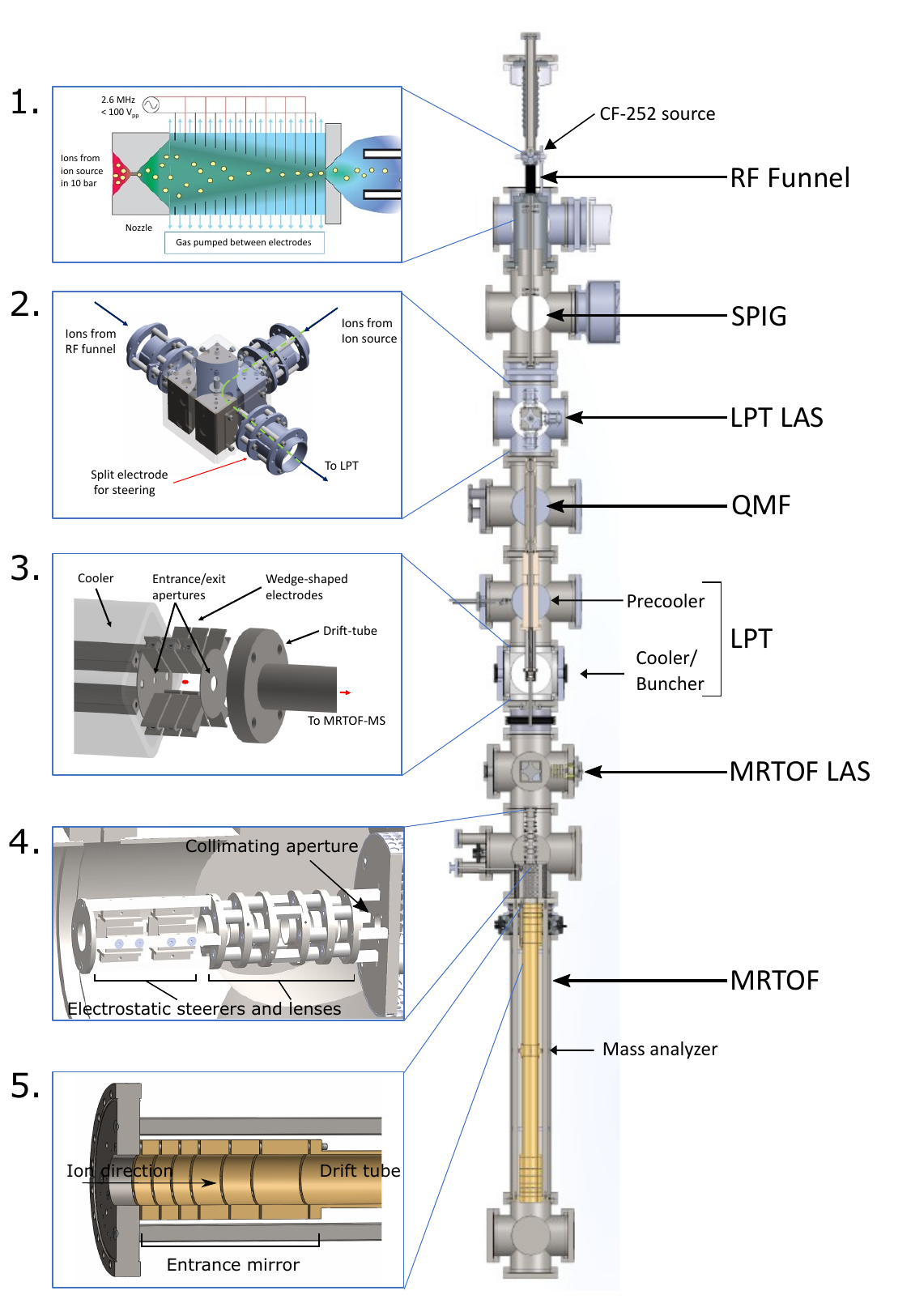}
    \caption{\textls[-15]{{CAD} 
 model rendering of the Ba-tagging setup for \bariumium\space extraction from GXe. Shown are the RF funnel, LPT, and MRTOF-MS going from top to bottom. Starting from pressures of up to 10 bar, ions pass through (1) the RF funnel and the SPIG to arrive in a high-vacuum environment in (2) the LPT LAS chamber. Lenses of this LAS allow the focusing and steering of ions, while the quadrupole bender allows selecting between ions from the LAS and the RF funnel. The QMF filters incoming ions based on their m/q ratio and guides the ions to the LPT. The precooler gradually cools ions, which are then trapped in (3) the cooler and buncher, which are used to cool and bunch the ions, respectively. After bunching, the ions are ejected to the MRTOF-MS, passing through the MRTOF-MS LAS bender, (4) ion optics, and a collimating aperture. In the MRTOF-MS, ions are reflected several times between (5) two sets of electrostatic mirrors before being detected using a CEM to obtain a time-of-flight spectrum.}}
    \label{fig:lpt_mrtof}
\end{figure}

The task of ion identification is divided between the LPT, where elemental identification of the Ba ion is performed using laser fluorescence spectroscopy, and the MRTOF-MS, where the mass of \bariumium\space is identified to be 136. Figure~\ref{fig:lpt_mrtof} shows the entire setup being developed for ion extraction from gas medium, including components used to facilitate their commissioning and testing. 

\subsubsection{Ion Sources}

While the $^{252}\text{Cf}$ source is currently used as an in-gas ion source, laser ablation ion sources (LAS) are positioned upstream of the LPT and the MRTOF-MS for ion transport optimization and to provide calibrant masses for the MRTOF-MS.

\subsubsection{Quadrupole Mass Filter (QMF)}
A QMF acts as an ion guide in its quadrupole design and the use of RF potentials but can additionally allow selective propagation of ions of a specific (or range of) mass-to-charge ratio(s). 
The QMF is positioned downstream from the RF funnel (see Figure~\ref{fig:lpt_mrtof}) to remove background ions such as the molecular species seen in Figure~\ref{fig:dan_plot} that exit the RF funnel. This will help improve the trapping efficiency of the LPT by reducing the number of ions being trapped. 

There are three sets of quadrupole electrodes in the QMF, each supplied with a DC-coupled RF potential. The main purpose of the first and third sets of quadrupole electrodes (referred to as ramp segments in~\cite{rasiwala2022development}) is to provide a delayed DC ramp and reduce potential losses due to fringing fields~\cite{dawson1971fringing}. The RF potential is supplied through a transformer, which in turn is supplied by an RF power amplifier and function generator, similar to the RF funnel. The RF frequency of the current operation is impedance matched for a \mbox{100 V$_{\text{pp}}$} sinusoidal signal at a frequency of 750 kHz. Initial tests performed using a cesium thermal source demonstrated a mass resolving power, m$/\Delta$m $\approx$ 100~\cite{rasiwala2022development}, of the QMF. This exceeds the design goal of m$/\Delta$m $>80$~\cite{lan2020linear}, which will allow the QMF to resolve most of the background ions aside from \xenon$^+$. Future tests will be performed by replacing the thermal source with a laser ablation ion source (LAS) using a multi-element target to produce ions over a large mass range. The ablated ions will be supplied to the QMF using a quadrupole bender and other focusing elements, as shown in Figure~\ref{fig:lpt_mrtof} (2). 

\subsubsection{Ion Cooling and Bunching (ICB)}
Filtered ions from the QMF are subsequently guided into the LPT, where they first pass through a precooler. The precooler is comprised of a quadrupole with a flow-limiting aperture that allows for differential pumping between the QMF and the cooler. Similarly, the cooler also has a quadrupole electrode structure, where ions are collisionally cooled and stored. Rather than using segmented electrodes, the quadrupole electrodes of the cooler are tapered along their length and are enclosed in a square tube~\cite{rasiwala2022development}. The axial potential gradient created due to this taper is used to store ions just upstream from the entrance aperture of the buncher. A constant flow of helium gas (25 sccm) is supplied by a mass flow controller to maintain a pressure of 0.1 mbar in the cooler for sufficiently cooling ions to kinetic energies of under 1 eV and storing them before being transferred to the buncher for identification. As shown in Figure~\ref{fig:lpt_mrtof} (3), the buncher consists of three sets of wedge-shaped quadrupoles for axial and radial confinement of transferred ions and aperture electrodes to control loading and unloading of the buncher. Ions from the cooler are loaded into the buncher during a 4 $\upmu$s transfer period by switching the bias voltage of the entrance aperture to allow ions into the buncher. At the end of the transfer period, the entrance aperture bias voltage is switched back to prevent ions from escaping the buncher while also preventing other ions from entering. In the buncher, the ions are cooled over a $\approx$10 ms period. Once trapped, the Ba ion will be identified using a combination of lasers to induce fluorescence that will then be detected by a charge-coupled device (CCD) camera~\cite{laser_spec,laser_spec_thesis}. This spectroscopic identification method will be implemented in the LPT in the future. Bunched ions are ejected from the buncher by applying a forward potential gradient using voltage switches, similar to the entrance aperture, and accelerated into a pulsed drift tube. The pulsed drift tube is switched to a different potential while the ions are traveling inside, which accelerates them further when they exit the tube before they are injected into the MRTOF-MS~\cite{rasiwala2022development}. The pulsed drift tube switching potentials define the kinetic energy of the ion bunch. 

\begin{figure}[H]
\begin{subfigure}{0.49\textwidth}
    \centering
    \includegraphics[width=\textwidth]{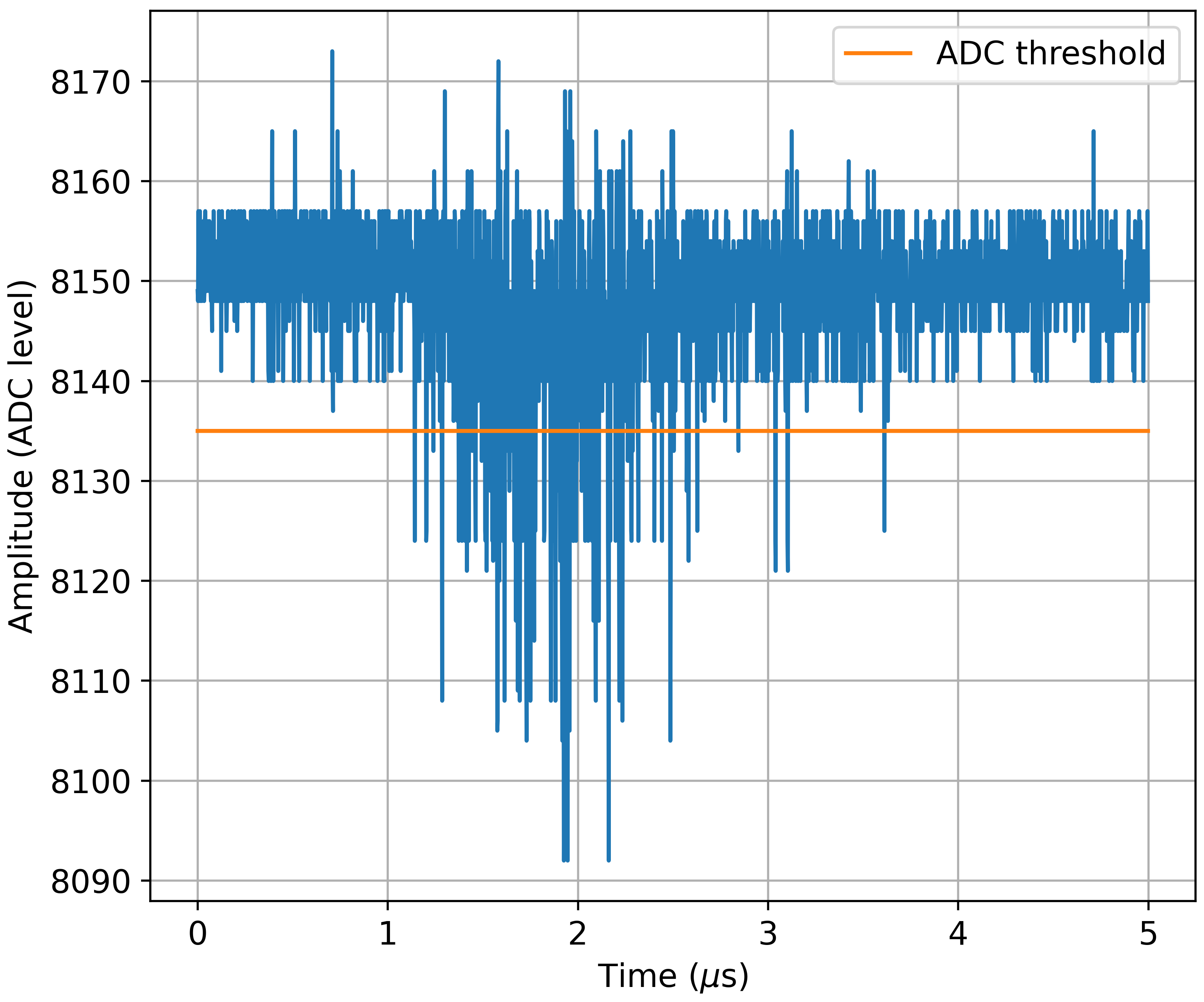}
    \caption{Typical waveform of ion bunch at CEM.\vspace{2mm}}
    \label{fig:ion_bunch_waveform}
\end{subfigure} 
\begin{subfigure}{0.49\textwidth}
 \centering
    \includegraphics[width=\textwidth]{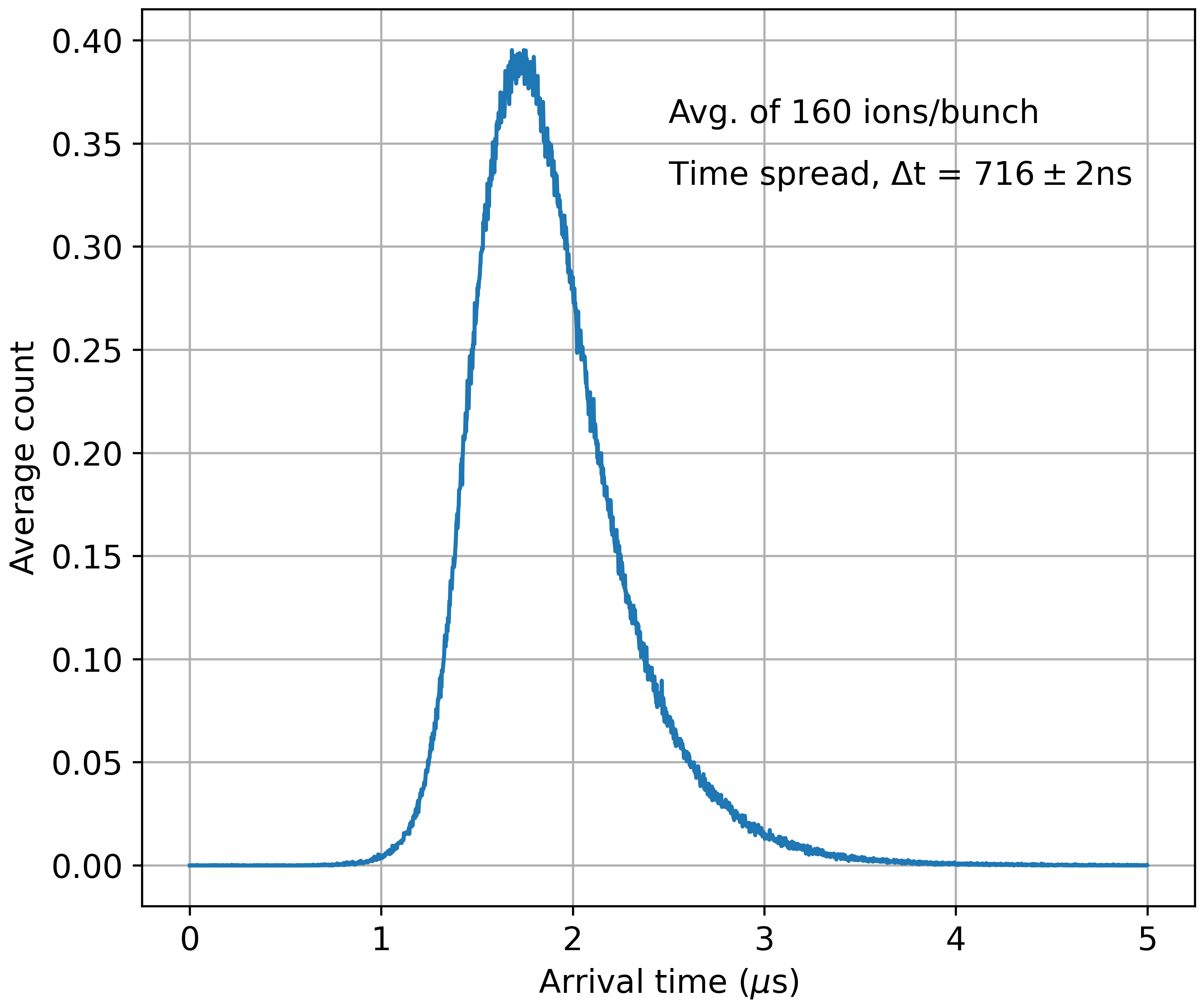}
    \caption{TOF spectrum of ion bunch. \vspace{2mm}}
    \label{fig:ion_bunch_dist}
\end{subfigure}      
\caption{{Bunched} 
 ion spectrum and waveform from run using bunched cesium ions. (\textbf{a}) Typical waveform of one ion bunch recorded using a CAEN DT5730. (\textbf{b}) Histogram of samples above the signal noise threshold, accumulated and averaged over 10,000 waveforms.}
\label{fig:ion_bunch}
\end{figure}

Ion bunching is currently being studied using a CEM located downstream from the buncher inside the MRTOF-MS LAS bender. Figure~\ref{fig:ion_bunch}a shows an example of waveform readout from the CEM during a test run using a CAEN DT5730 digitizer and a continuous Cs ion beam from a thermal ion source installed upstream from the QMF. Samples that are below the ADC threshold\endnote{Threshold is set as the minimum \mbox{(data without ions) $-$ 5.} Threshold in Figure~\ref{fig:ion_bunch}a = 8140 $-$ 5 = 8135.} are recorded as ion hits on the CEM. The number of ions detected in a given ion hit is then inferred from the analog-to-digital converter (ADC) amplitude. Figure~\ref{fig:ion_bunch}b shows the time distribution of ions averaged over 10,000 such waveforms. Operational parameters such as buffer gas pressure, cycle time for trapping and ejection, and buncher potentials are currently being optimized to reduce the time spread of the ion bunch since it has a direct impact on the mass resolving power of the MRTOF-MS~\cite{murray2023design}. 
Additionally, the thermal ion source will be replaced with a multi-element target and used to study the operation of the LPT and the MRTOF-MS over a large mass range. This will also allow the detection of molecular ions such as $[\text{XeXe}]^+$ and $[\text{BaXe}_{\text{n}}]^+$ that were observed in the previous RF funnel study~\cite{fudenberg2018improved} and a mobility study of Ba ions in GXe~\cite{medina2014mobility}, respectively. Aside from $[\text{BaXe}_{\text{n}}]^+$ detection, a combination of RF heating and collision-induced dissociation using the helium buffer gas in the LPT may also be used to separate the Ba ion from Xe.
\vspace{-4pt}

\subsubsection{Multiple Reflection Time-of-Flight Mass Spectrometer (MRTOF-MS)}

The MRTOF-MS accepts bunched ions from the LPT and performs mass spectrometry to identify \bariumium\space from a background of $^{136}\text{Xe}^+$ as well as perform systematic studies. In place of the bunched ions from the LPT, for commissioning and testing, a LAS is used~\cite{murray2022characterization}. A 349 nm Spectra Physics laser is focused onto a copper target to ablate material and in the process produce ions. Ions produced by the LAS travel through a quadrupole bender and other ion optics (see Figure~\ref{fig:lpt_mrtof} (4)) to reach the MRTOF-MS. The MRTOF-MS operates by reflecting ions several times between two sets of electrostatic mirrors, shown \mbox{in Figure~\ref{fig:lpt_mrtof} (5).} Since ions with the same kinetic energy but different mass-to-charge ratios travel at different velocities, multiple reflections greatly increase the difference in their time of flight and thus resolve the different mass peaks. Electrodes of the electrostatic mirrors are biased using two-state and three-state high-voltage switches\endnote{High-voltage three-state switches (\url{http://madcow-elec.com}, {accessed on 14 November 2024).} 
} to change potentials during the first reflection after the injection of the ion bunch and the last reflection before detection.  

The MRTOF-MS commissioning was completed in 2022, and a mass resolving power of 20,000 was demonstrated~\cite{murray2023design}. However, this was achieved using ions from the LAS that are not bunched but created in ablation with laser pulses. Simulation studies of the MRTOF-MS using bunched ions from the LPT showed a mass resolving power over 100,000 can be achieved, which is sufficient to resolve $^{136}\text{Ba}^+$ and $^{136}\text{Xe}^+$. Thus, following the optimization of the buncher, bunched ions will be injected into the MRTOF-MS to demonstrate its resolving power.

\section{In-LXe Ion Source for Testing Ba Ion Extraction Techniques}\label{sec:ion_source}

A controlled and calibrated Ba-ion source is required for the definitive demonstration, final optimization, and quantification of Ba extraction from LXe with either the cryoprobe or the capillary method. Research and development (R\&D) efforts are currently ongoing to develop an accelerator-driven, radioactive ion source using rare isotope beams (RIB) from the Isotope Separator and Accelerator (ISAC) facility~\cite{TRIUMF_ISAC_Ball_2016} at TRIUMF, \mbox{{Vancouver, BC,} 
 Canada.}
An apparatus previously employed at Stanford University, {Stanford, CA,}
 USA, for Ba-tagging developments using an internal, spontaneous fission $^{252}$Cf source and a deposition substrate for the Resonance Ionization mass Spectroscopy (RIS) technique~\cite{twelker2014apparatus,Twelker:2014bfy,Kravitz:2017imi} is being recommissioned for a proof-of-principle measurement.
A beam of radioactive $^{139}$Cs ions ($T_{1/2}= 9.27(5)$ m~\cite{nubase2020}) will be injected through a Be window and stopped in LXe, where the decay-daughter $^{139}$Ba ions will be tagged by collecting and extracting them electrostatically and detecting them using $\gamma$~spectroscopy. 
Once the  concept is tested, further developments will be conducted to make future ion sources that will be suitable for the cryoprobe and the capillary extraction techniques.
For a final demonstration and determination of Ba-tagging efficiency, injecting $^{136}$Cs ($T_{1/2}= 13.01(5)$~d~\cite{nubase2020}) with $Q_{\beta} \approx$ 2.5~MeV~\cite{MCCUTCHAN2018331} and tagging its decay daughter $^{136}$Ba is also being considered.

\begin{figure}[H]
\hspace{-5pt}
    \includegraphics[width=\textwidth]{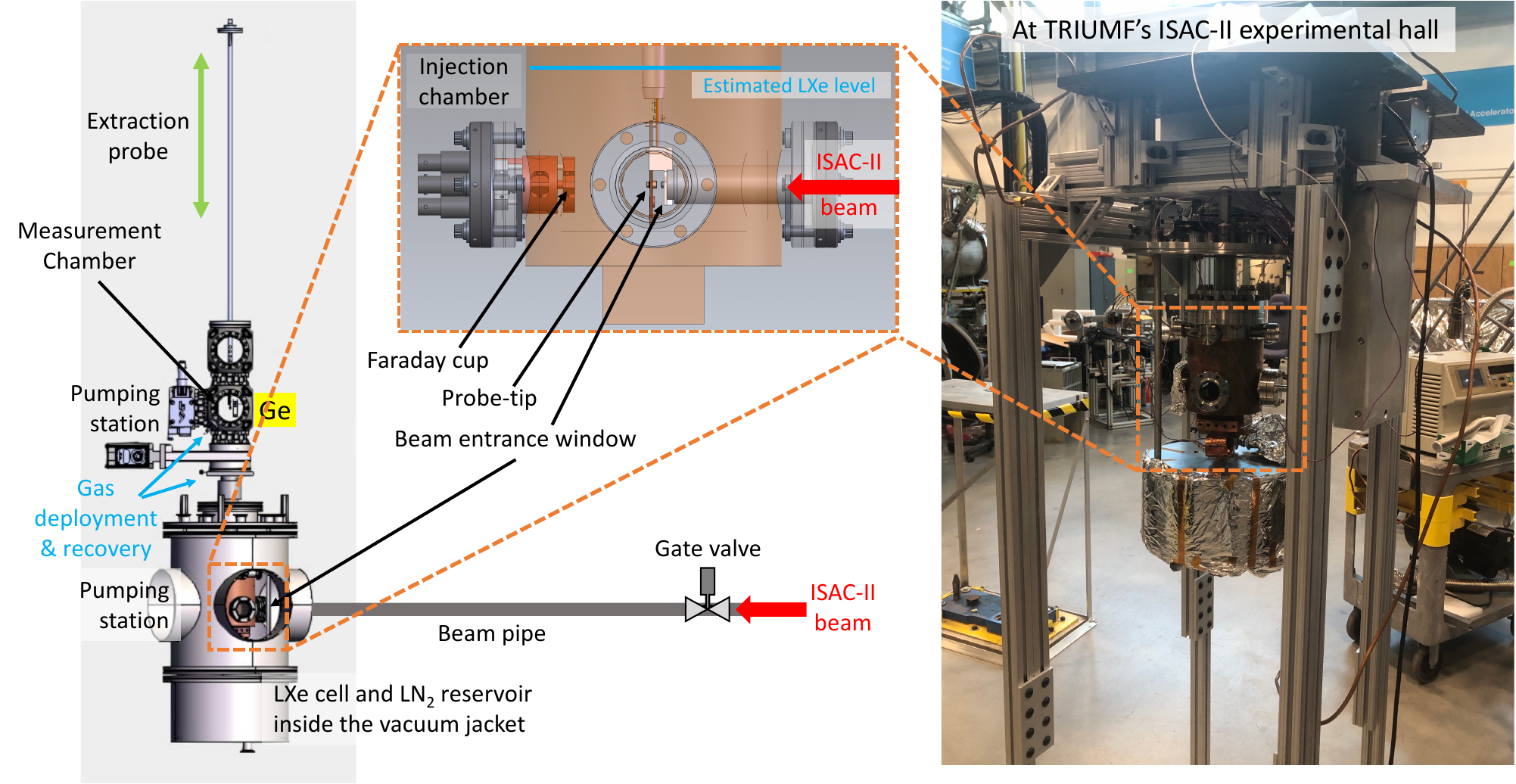}
    \caption{Ion source for Ba extraction tests at TRIUMF. {(\textbf{Left})} 
 An engineering rendering of the apparatus showing the IC enclosed by a vacuum jacket, the MC, and the extraction probe. \mbox{{(\textbf{Inset})} 
~Zoomed} view of the engineering rendering of the IC, labeling the beam entrance window, the probe tip (or electroprobe), and the Faraday cup, and showing the estimated LXe fill level in a light blue line. \mbox{{(\textbf{Right})}
 Photograph} of the setup being assembled at TRIUMF's ISAC-II experimental hall. See text \mbox{for details.}}
    \label{fig:TRIUMF_BaTag_sys}
\end{figure}

An engineering rendering of the apparatus, along with a picture of it being assembled at TRIUMF's ISAC-II experimental facility, is shown in Figure~\ref{fig:TRIUMF_BaTag_sys}. The apparatus includes three main parts: an Injection Chamber (IC), a Measurement Chamber (MC), and an extraction probe that can be moved between the IC and the MC. 
The IC consists of an LXe cell of volume 1~L with 1.9~cm thick copper walls, thermally coupled to a liquid nitrogen (LN$_{2}$) reservoir with copper heat-transfer straps, and is suspended in a vacuum chamber for thermal insulation. The high thermal conductivity of the copper straps means the LN$_{2}$-enabled cooling system over-cools the IC. 
The copper straps are equipped with resistive heaters controlled by a PID controller using temperatures measured with PT100 probes. Thus the cell is maintained at $\sim$165 K, and the Xe is kept in its liquid state.
Four equally spaced DN40 ports are placed radially around the LXe cell. These ports will be fitted with a beam entrance Be-window mount, a diagnostic Faraday cup detector opposite the Be window, and two side-view ports. 
The 25~$\upmu$m thick beam entrance Be window will be metal diffusion bonded on the front face of the nozzle of the mount to separate the LXe inside the cell from the upstream beamline vacuum.
The nozzle protrudes $\sim$8~cm into the LXe.
This is shown on the inset of Figure~\ref{fig:TRIUMF_BaTag_sys}.
The MC is a small chamber above the IC with a high-purity Ge detector (HPGe) placed at a view port for $\gamma$ spectroscopy.
The extraction probe, controlled by a linear actuator with a stepper motor, consists of a flat rod with the removable copper target at its tip (also referred to as an electroprobe; see inset of Figure~\ref{fig:TRIUMF_BaTag_sys}).
Radioactive ions will be collected by biasing the target while it is placed in LXe in front of the beam entrance window in the IC. For identification, the target will be positioned in front of the HPGe. The setup will be connected to a beamline in the high-energy RIB facility ISAC-II (called the SEBT-I beamline) through a beam pipe and a gate valve.

The LXe will be contained at a pressure of 1~bar inside the IC, which requires around 3.1~kg of gaseous Xe (GXe).
Given the cost associated with procuring this required quantity of Xe, a closed gas handling system (GHS) is being developed for deploying and recovering GXe with minimal to no loss.
The gas will be deployed from the GHS into the apparatus through two ports on either side of a gate valve separating the IC and the MC (labeled as ``Gas deployment \& recovery'' in Figure~\ref{fig:TRIUMF_BaTag_sys}) from a pre-filled gas bottle through a cold getter purifier to ensure impurities are removed from the gas. 
The recovery will occur through cryopumping GXe into a second gas bottle, which will act as the supply bottle for the next deployment cycle.

The radioactive ions will be produced at TRIUMF's ISAC facility~\cite{TRIUMF_ISAC_Ball_2016} by impinging a $\sim$480~MeV proton beam of $\sim$40~$\upmu$A from the TRIUMF cyclotron~\cite{TRIUMF_cyclotron} onto a uranium-carbide (UC$_{x}$) target by fragmentation, spallation, and fission reactions. The $^{139}$Cs ions will be surface ionized before being extracted and mass separated at ISAC's mass separator~\cite{ISAC_magnet_separator_BRICAULT200249}. The mass-separated beam will pass through the low-energy beam transport (LEBT)~\cite{LEBT_SEN201697} electrostatic beam line and then through a radio-frequency quadrupole (RFQ)~\cite{laxdal_2014_isac2}, where its longitudinal emittance will be lowered, before being directed towards ISAC-II. Next, the beam will pass through a charge state booster~\cite{charge_state_booster_Adegun_2022} before being reaccelerated using a drift tube linac (DTL) and superconducting linear accelerator (SC-linac)~\cite{laxdal_2014_isac2} and delivered to the experiment at the SEBT-I beamline with an intensity of $\sim$10$^{5}$~particles per second~(pps) at an energy of up to 10 MeV$/$u.

\begin{figure}[H]
\hspace{-5pt}
    \includegraphics[width=\textwidth]{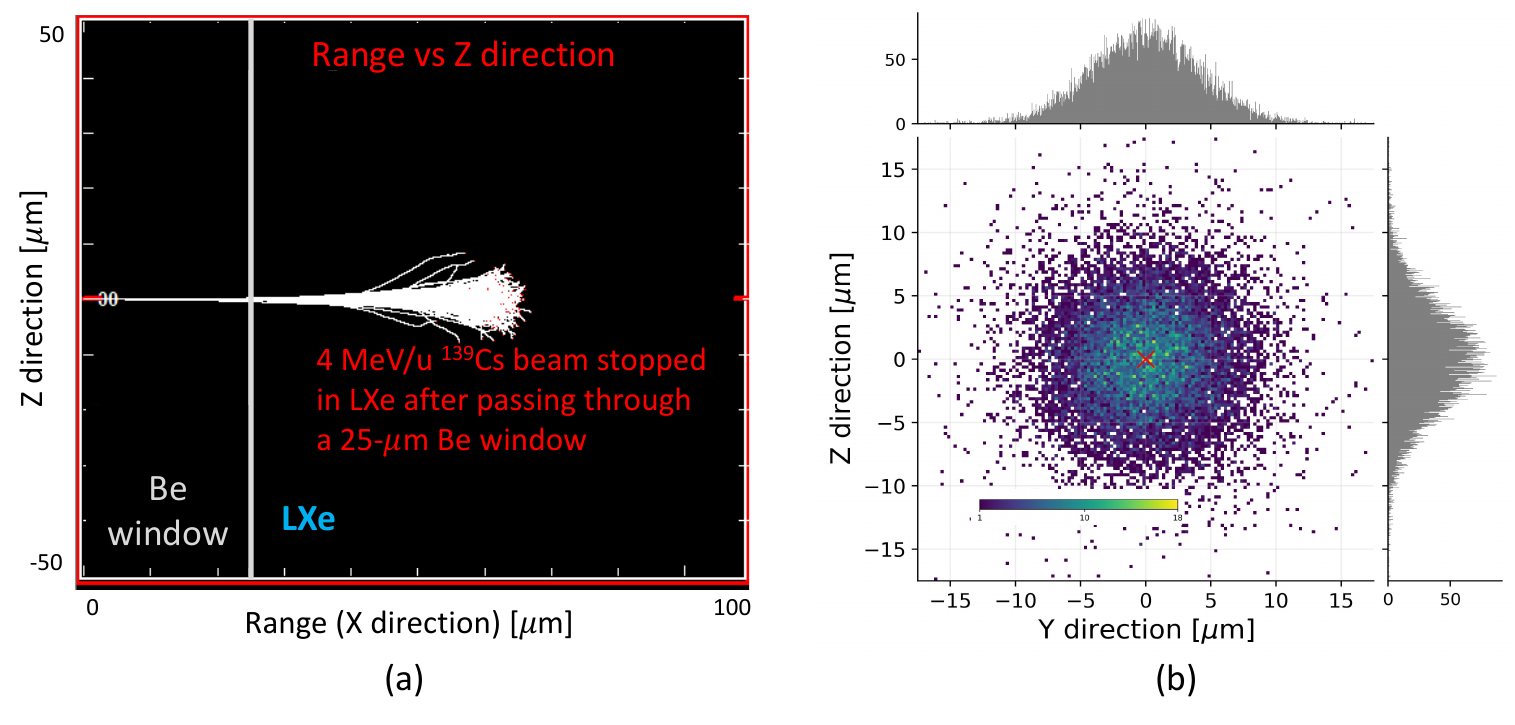}
    \caption{{Simulations} 
 of ion implantation inside LXe, at 1~bar pressure, having a density of 3.1~kg$/$L, after passing through a 25~{$\upmu$m} Be window, with a density of 1.845~kg$/$L, using TRIM. (\textbf{a}) TRIM simulations showing {$^{139}$Cs} 
 ions with energy of $4$~MeV$/$u being stopped in LXe. The white vertical line represents the boundary between the Be window and the LXe medium. (\textbf{b}) Lateral (in Y-direction) and longitudinal (in Z-direction) straggling of a beam of maximum possible energy of 10~MeV$/$u. The red cross around (0, 0) marks the hypothetical alignment point of the probe tip.
    \label{fig:TRIUMF_trim}}
\end{figure}

At the start of the experiment, the probe will be inserted into the LXe cell using the linear actuator, such that the collection tip lines up with the beam entrance window, as shown on the inset of Figure~\ref{fig:TRIUMF_BaTag_sys}.
GXe will be introduced into the system from the GHS and liquified in situ until the LXe reaches a set level (blue line on the inset of Figure~\ref{fig:TRIUMF_BaTag_sys}), sufficiently submerging the probe tip. Note that the volume of the IC above the blue line and the MC will be filled with Xe vapor. 
An upstream beamline valve will then be opened, and a beam of $^{139}$Cs, with some contaminant $^{139}$Ba ions, with energies between 2 and 10~MeV$/$u, will be injected through the Be window into the LXe.
Monte Carlo simulations of the ions' trajectories using Transport of Ions in Matter (TRIM)~\cite{ziegler2011stopping} with a beam of $^{139}$Cs in LXe, at 1~bar pressure and having a density of 3.1~kg$/$L, with one such energy (4~MeV$/$u) and through a 25~$\upmu$m Be window, with a density of 1.845~kg$/$L, show an implantation depth of $\sim$35~$\upmu$m from the window (Figure~\ref{fig:TRIUMF_trim}a). The straggling of the ions with maximum possible energy of 10~MeV$/$u, obtained from Stopping and Range of Ions in Matter (SRIM)~\cite{ziegler2011stopping}, is shown in Figure~\ref{fig:TRIUMF_trim}b.
Thus, upon stopping, the ions will be within a few millimeters of the probe tip. After a few half-lives of $^{139}$Cs, the beam implantation will be stopped and a negative voltage will be applied on the probe tip to induce an electric field in the LXe medium and initiate electrostatic collection.

\begin{figure}[H]
\hspace{-2pt}
    \includegraphics[width=\textwidth]{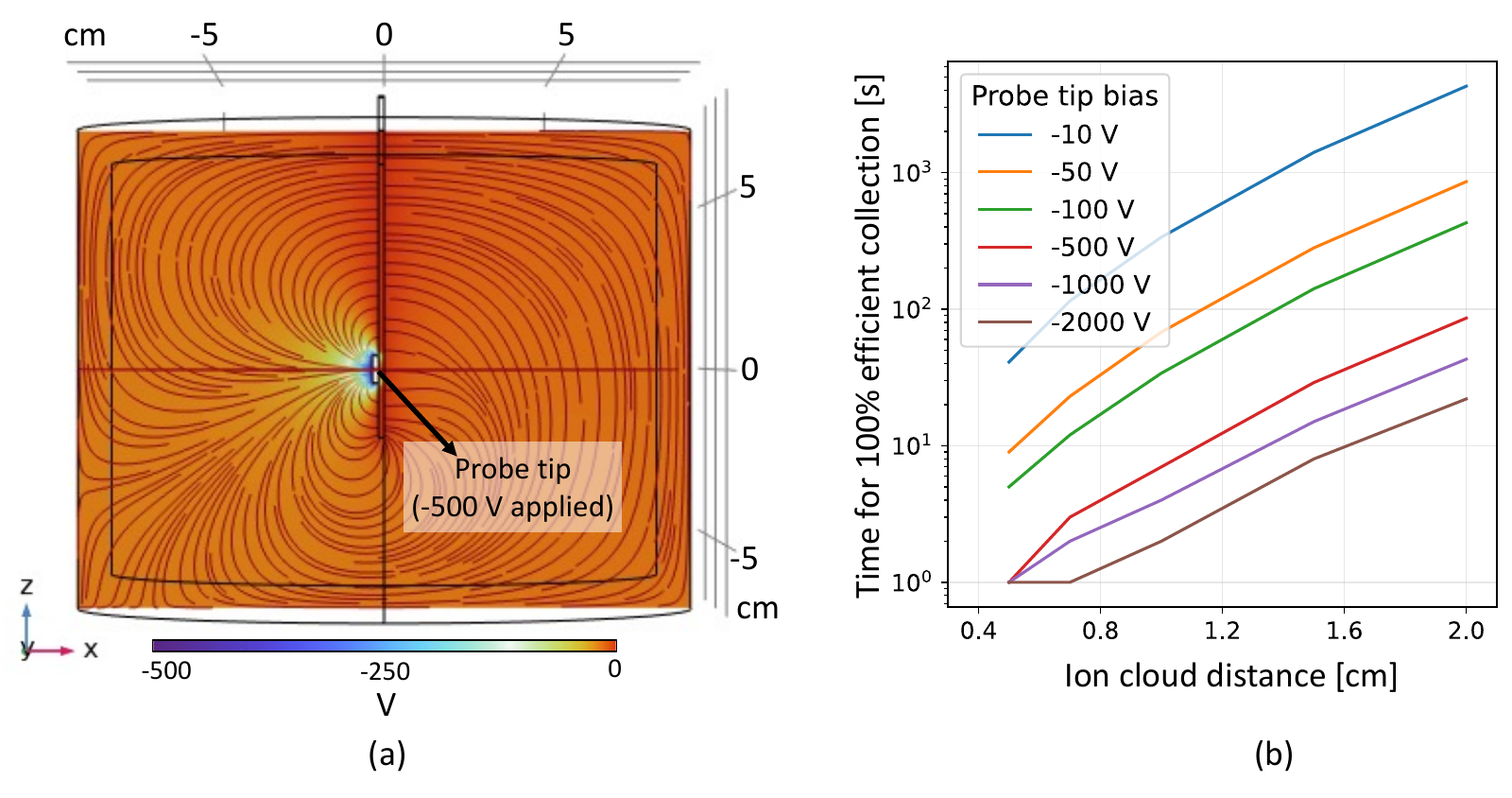}
  
\vspace{-7pt}  \caption{{Simulations} 
 of ion collection in LXe using COMSOL Multiphysics. The ions enter the volume from the left side. (\textbf{a}) Electric potential in LXe generated by the application of $-500$ V on the probe tip, with the field lines overlaid on top of that heat map.
    (\textbf{b}) The time required for 100\% efficient collection for ions starting from different positions for different probe biases.
    \label{fig:TRIUMF_comsol}}
\end{figure}

The collection of the ions is simulated with {COMSOL Multiphysics\texttrademark,} 
 using the AC$/$DC, the computational fluid dynamics (CFD), and the Particle Tracing for Fluid Flow modules. The simulation consists of a simplified copper vessel filled with LXe at 1~bar and 165~K and an electroprobe, as shown in Figure~\ref{fig:TRIUMF_BaTag_sys} inset and Figure~\ref{fig:TRIUMF_comsol}a. 
For this 3D geometry, COMSOL is used to solve electrostatics and fluid dynamics within the vessel's environment, as well as particle tracing for the singly charged ions released within the LXe. 
The electrostatics model starts with the negative voltage being applied to the electroprobe while keeping the rest of the setup grounded. The LXe is taken as an ideal dielectric medium, which can be polarized by the electric field generated from the probe. The simulation then computes the induced electric field lines in the LXe 
and the electric force experienced by the positive Ba ions by solving a modified version of Gauss' law for dielectric media and Faraday's law of electrostatics, since the field is irrotational. This is shown in Figure~\ref{fig:TRIUMF_comsol}a.

Each COMSOL run starts with inputting lateral and longitudinal straggling information, as adapted from SRIM for a beam of maximum possible energy (10~MeV$/$u) (Figure~\ref{fig:TRIUMF_trim}b), the ions' mass, diameter, and charge. Next, the ions are released from different initial $X$ positions between the window and the probe tip. With the initial conditions, the simulation solves ordinary differential equations for the velocity of each particle as it travels through the LXe medium. It is assumed that the ions do not displace the fluid in which they are submerged. The motion of the Ba ions in LXe is dictated by the electric and the drag forces, as solved by the electrostatics and the fluid-dynamics model.
The collection efficiency is calculated by comparing how many initially released ions are present on the boundary of the probe tip surface after a given amount of time.
Figure~\ref{fig:TRIUMF_comsol}b shows the amount of time required to collect all of the ions onto the probe (100\% collection efficiency) from different starting positions for various probe biases. 
The results follow an expected trend of higher applied voltage leading to ions feeling a stronger pull and hence getting collected faster. 
These simulation results provide a reference for the experimental parameters, namely probe bias and beam implantation and collection time, which will then be optimized and finalized from initial runs during experimental campaigns.
Ions may be neutralized by contaminants in the LXe and free charges generated when stopping the beam. In parallel, the $Q_{\beta}$ of the decays could again force the neutral atoms to be ionized to the $1+$ or $2+$ states. These competing effects will give a non-zero probability of the final product being an atom, not an ion. We will be determining this neutralization fraction during the commissioning campaigns. 

\vspace{-8pt}

\begin{figure}[H]
    \includegraphics[width=\textwidth]{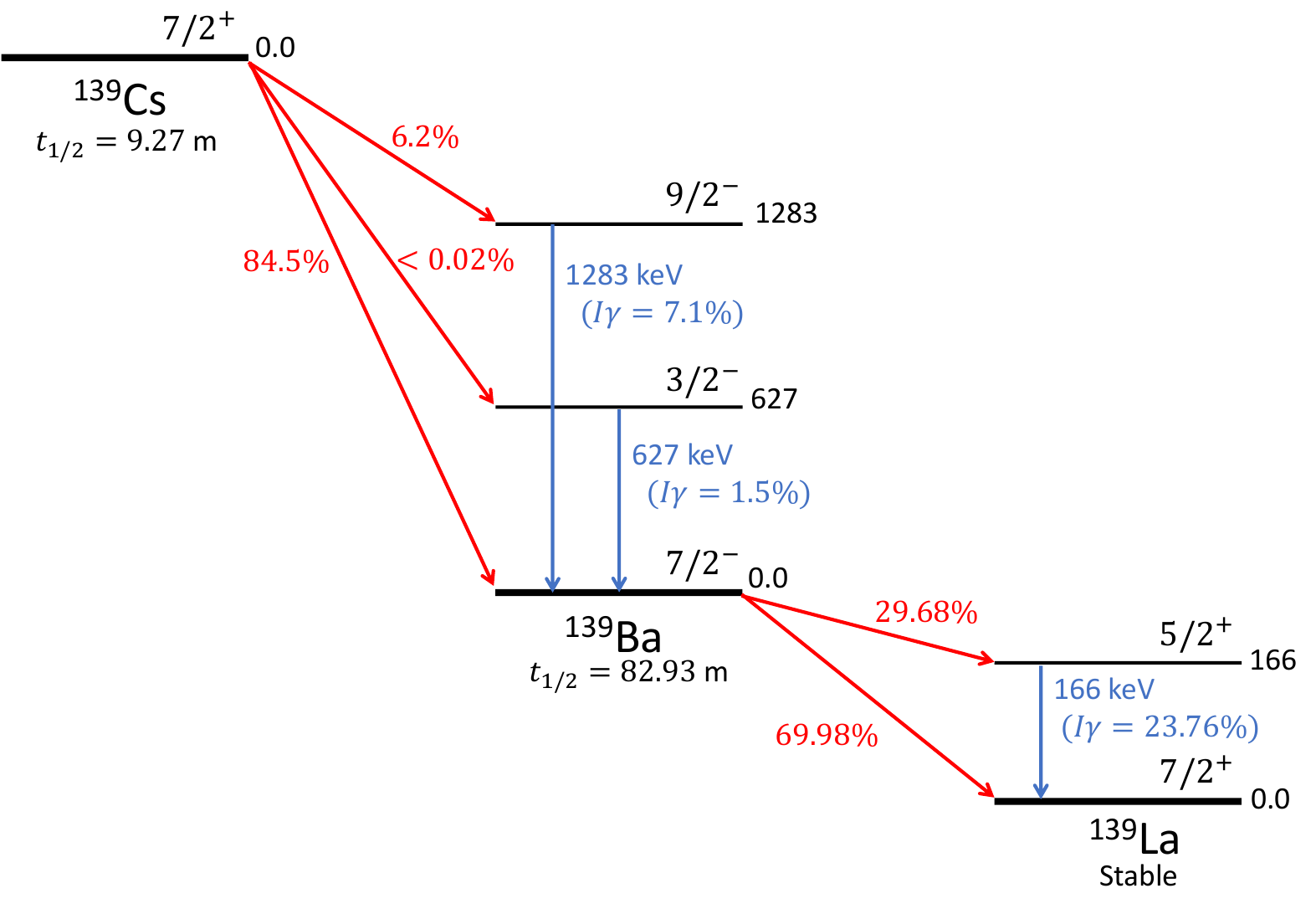}

\vspace{-6pt}\caption{Decay scheme for $^{139}$Cs $\rightarrow$ $^{139}$Ba $\rightarrow$ $^{139}$La showing the $\beta$-decay (in solid red arrows) and $\gamma$-decay (in solid blue arrows) branches for the relevant levels. The energies are in keV, and the levels are not to scale. Data taken from Ref.~\cite{JOSHI20161}.
\label{fig:TRIUMF_levels}}
\end{figure}

After passage of an optimized collection time, the probe will be retracted using the actuator such that the tip moves from the IC to the MC and is placed in front of the HPGe for identification using $\gamma$ spectroscopy.
The decay scheme for $^{139}$Cs $\rightarrow$ $^{139}$Ba $\rightarrow$ $^{139}$La~\cite{JOSHI20161} is shown in Figure~\ref{fig:TRIUMF_levels}.
The extracted sample would predominantly consist of ions of $^{139}$Ba ($T_{1/2}= 82.93(9)$ m~\cite{nubase2020}), which will decay to $^{139}$La, populating its first excited level \mbox{166~keV ($J^{\pi}=5/2^{+}$).}
Therefore, the primary goal is to detect the 166~keV $\gamma$ photons emitted from internal transitions from this 166~keV level to its ground state ($J^{\pi}=7/2^{+}$)~\cite{JOSHI20161}.
In addition, some $^{139}$Cs ions could also get extracted, which will decay to $^{139}$Ba, populating its 1283~keV level ($J^{\pi}=9/2^{-}$) and 627~keV level ($J^{\pi}=3/2^{-}$).
As a result, we will also aim to detect the emitted $\gamma$ photons from internal transitions from these levels to the ground state ($J^{\pi}=7/2^{-}$). This will prove the extraction of Ba ions, either from the injected beam or from the decay of the injected Cs ions, which can be determined from the preestablished beam composition and intensity. Any unwanted extracted species, like isobaric contaminants, present in the beam can be differentiated by their distinct $\gamma$ signatures.

\begin{figure}[H]
\hspace{1pt}
    \includegraphics[clip,width=.9\textwidth]{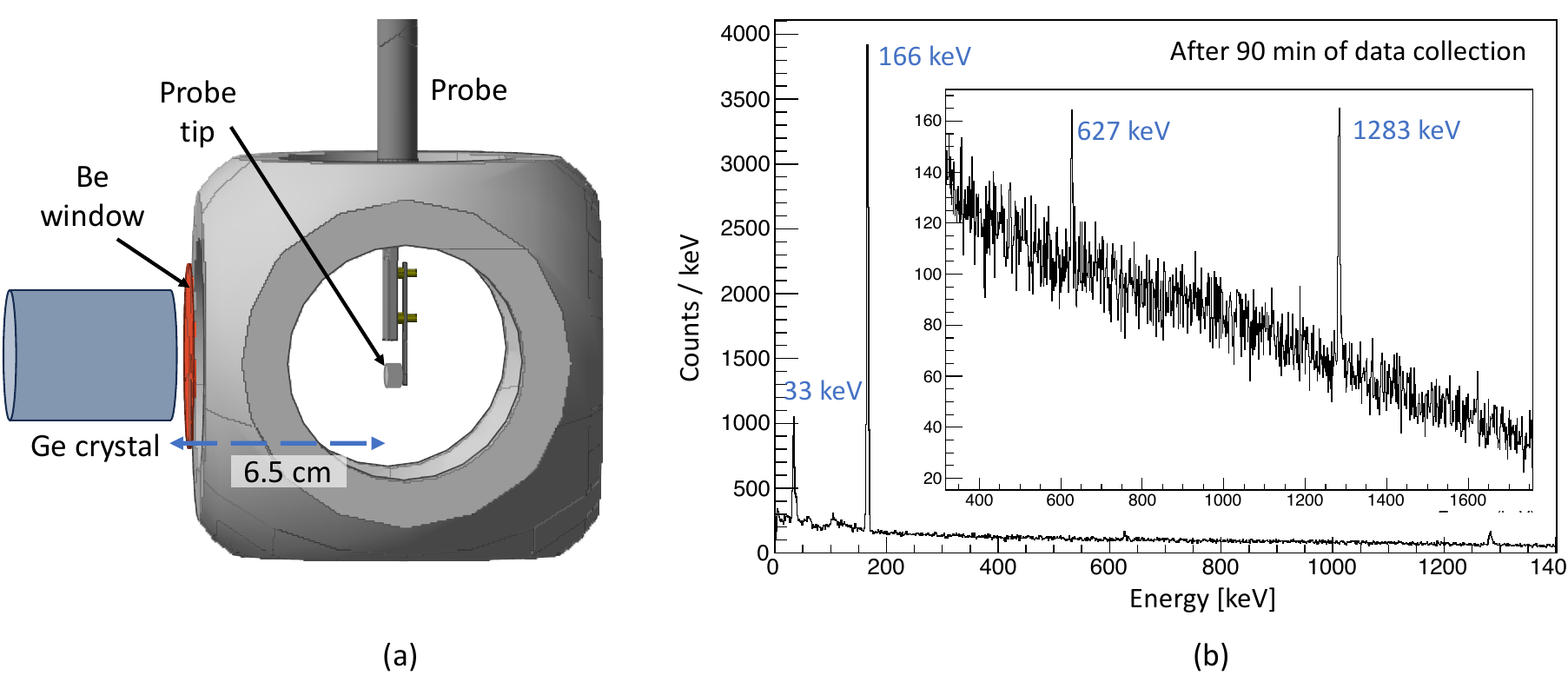}
  \vspace{-6pt}  \caption{{Geant4} 
 simulations: (\textbf{a}) The model with the chamber hosting the probe tip, the Be window, and the Ge crystal of the HPGe detector. (\textbf{b}) Photon spectrum acquired over 90 min from a deposit containing $10^{7}$ ions of $^{139}$Cs and $^{139}$Ba in a ratio of 1$:$1. The 166~keV peak from internal transitions in $^{139}$La is labelled. Also shown is a 33~keV X-ray peak. The inset is zoomed in on the energy range between $\sim$310~keV and $\sim$1760~keV to show the 627~keV and 1283~keV peaks from internal transitions in $^{139}$Ba.
    \label{fig:TRIUMF_geant}}
\end{figure}

Simulations were conducted in Geant4~\cite{AGOSTINELLI2003250_Geant4}, where the experimental apparatus was modeled.
The model consisted of three main parts: a probe tip implanted with a radioactive ion source; a GXe-filled chamber having the same dimensions as the MC in Figure~\ref{fig:TRIUMF_BaTag_sys} with a Be window of thickness 0.25 mm on one flange; and an HPGe, with a Ge crystal placed 6.5~cm from the probe tip next to the Be window.
The Ge crystal used was 6.68~cm in length and 5.76~cm in diameter, obtained from the manual of the HPGe, and had a 2~mm thick inner copper shield and a 2~mm thick outer lead shield around it. This is shown \mbox{in Figure~\ref{fig:TRIUMF_geant}a.}
The experimental hall's photon background, as well as detector efficiency and energy resolution, were obtained from past $\gamma$ spectroscopy experiments using the same HPGe detector. 
Figure~\ref{fig:TRIUMF_geant}b shows a simulated $\gamma$ spectrum using a source containing \mbox{$10^{7}$ ions} of $^{139}$Cs and $^{139}$Ba in a ratio of 1$:$1 integrated over 90 min of data acquisition.
The HPGe will be recommissioned to obtain the current background, detector efficiency, and energy resolution and will be used for initial tests of the ion source.

\vspace{-12pt}

Initial tests will be conducted using gas media, including a commissioning run using gaseous argon (GAr), followed by a run using GXe. 
Simulations have been repeated for GAr and GXe at 1 bar pressure with densities of 1.784 g$/$L and 5.86 g$/$L. The major differences between the gas and liquid media are the ions' ranges being $\sim$100 times longer and the collection times being $\sim$10 times faster in gas media.
To accommodate these, a spacer will be used to create an additional distance of $\approx$10 cm between the probe tip and the beam entrance window.
A proposal for an experiment using GAr has been approved with high priority by the Nuclear Physics Experiments Evaluation Committee at TRIUMF for 9 RIB shifts.
The campaigns using gas media will be used to optimize the GHS, beam intensities, implantation and ion collection durations, and detection procedures, including the HPGe-detector position and the necessary data-acquisition times, before the experiment will be repeated with LXe. 
From a known beam intensity, beam composition, and number of detected $\gamma$ photons, one can determine the extraction efficiency. Detection of the 166~keV $\gamma$ photons from ions extracted from LXe will demonstrate that an accelerator-driven ion source is well suited for future testing and perfecting Ba ion extraction using the capillary tube or the cryoprobe techniques.

\section{Conclusions and Outlook}
The Ba-tagging scheme outlined in this paper is a comprehensive approach to extract ions from LXe, separate the ions from background gas, trap and probe the ions with laser spectroscopy, and then measure their mass using an MRTOF-MS. This involves preserving a single ion through several stages of an apparatus, over a change of eleven orders of magnitude in background pressure. The capillary extraction probe has been fabricated, and currently work is being conducted to demonstrate ion extraction, first in gas, then in LXe. The RF-driven ion separation stages have been refurbished and will soon demonstrate high-efficiency ion guiding. The MRTOF-MS has demonstrated a mass resolving power of 20,000 and is anticipated to exceed 100,000 in the near future. The online Ba-ion source at TRIUMF will soon be taking beam, with the first step to demonstrate ion collection and extraction from GAr, followed by similar experiments in GXe and finally LXe. Once all of the stages have been successfully demonstrated individually, they will be combined into a single apparatus for demonstration of the entire technique using the online LXe ion source. Lastly, a ton scale demonstrator is considered for final proof of the technique. Ba-tagging is a powerful tool that has the potential to eliminate all non-$\beta \beta$ radioactive backgrounds in $0\nu\beta\beta$ searches in Xe.



\vspace{6pt} 




\authorcontributions{
Conceptualization, T.B., T.D., J.D., W.F., R.G., G.G., A.A.K., K.G.L., A.L., V.V. and L.Y.; Data curation, D.R., R.C., H.R., I.C., M.C. and K.M.; Formal analysis, D.R., R.C., H.R., A.V.B., C.C., M.C., R.E., D.F., K.M., R.S. and V.V.; Funding acquisition, T.B., T.D., J.D., W.F., R.G., G.G., A.A.K., A.L. and L.Y.; Investigation, D.R., R.C., H.R., L.B., A.V.B., T.B., I.C., C.C., M.C., T.D., J.D., R.E., D.F., R.G., G.G., A.A.K., A.L., M.M.-P., K.M., R.S. and V.V.; Methodology, D.R., R.C., H.R., L.B., A.V.B., T.B., I.C., C.C., M.C., T.D., J.D., R.E., W.F., D.F., R.G., G.G., A.I., A.A.K., K.G.L., A.L., M.M.-P., K.M., K.O., R.R., R.S., X.S., J.S. and L.Y.; Project administration, T.B., T.D., J.D., W.F., R.G., G.G., A.A.K., A.L. and L.Y.; Software, D.R., R.C., H.R., L.B., C.C., M.C., R.E., R.S. and V.V.; Supervision, D.R., R.C., H.R., T.B., C.C., T.D., J.D., W.F., R.G., G.G., A.A.K., K.G.L., A.L., Z.L., K.M., V.V. and L.Y.; Validation, T.B.; Visualization, D.R., R.C., H.R., L.B., T.B., I.C., M.C., R.E. and R.S.; Writing---original draft, D.R., R.C. and H.R.; Writing---review and editing, L.B., A.V.B., T.B., I.C., C.C., M.C., T.D., J.D., R.E., W.F., D.F., R.G., G.G., A.I., A.A.K., K.G.L., A.L., Z.L., M.M.-P., K.M., K.O., R.R., R.S., X.S., J.S., V.V. and L.Y. All authors have read and agreed to the published version of the manuscript.}

\funding{{This} 
 work has been supported by the Natural Sciences and Engineering Research Council of Canada {(NSERC) under Grant No. SAPPJ-2019-00058 and SAPPJ-2024-00034,} the Canada Foundation for Innovation {(CFI)} through the John R. Evans Leaders Fund {under Fund No. 35700}, and the Canada First Research Excellence Fund (CFREF) through the Arthur B. McDonald Canadian Astroparticle Physics Research Institute.
In the USA, support has been provided by the National Science Foundation (NSF) {under Grant No. 2011948}.
The ion funnel device and parts of the ion transport system were funded by NSF grant PHY-0918469 at Stanford University.
{The nEXO experiment is supported in Canada through NSERC Grant No. SAPPJ-2022-00021 and CFI Fund No. 43140 and 39881.}
}

\dataavailability{{The data presented in this paper are available on request from the corresponding authors.} 
}

\acknowledgments{{We} 
 would like to thank the nEXO collaboration. We thank M. Good, T. Koffas, S. Kravitz, P. Lu, M. P. Reiter, B. Schultz, P. Schury, {A. Stockton} and K. Twelker. D.R. would like to also thank the TITAN group (TRIUMF) and the Detectors group (TRIUMF).}


\conflictsofinterest{{The authors declare no conflict of interest. The funders had no role in the design of the study; in the collection, analyses, or interpretation of data; in the writing of the manuscript; or in the decision to publish the results.} 

} 



\newpage
\abbreviations{Abbreviations}{
The following abbreviations are used in this manuscript:\\

\noindent 
\begin{tabular}{@{}ll}
$0\nu\beta\beta$ & Neutrinoless double beta decay\\
$2\nu\beta\beta$ & Two Neutrino double beta decay\\
$\beta\beta$ & Double beta decay\\
CCD & Charge coupled device\\
CF & ConFlat\\
DTL & Drift tube linac\\
EXO-200 & Enriched Xenon Observatory\\
GHS & Gas handling system\\
GXe & Gaseous Xe\\
HPGe & High-purity germanium detector\\
ISAC & Isotope Seprator and Accelerator facility at TRIUMF\\
IC & Injection chamber\\
LAS & Laser ablation ion source\\
LEBT & Low energy beam transport\\
LPT & Linear Paul ion trap\\
LXe & Liquid Xe\\
MC & Measurement chamber\\
MRTOF-MS & Multi-reflection time-of-flight mass spectrometer\\
PID & Proportional integral derivative\\
PMT & Photomultiplier tube\\
pps & Particles-per-second\\
QMF & Quadrupole mass filter\\
RIB & Rare isotope beam\\
RIS & Resonance ionization mass spectroscopy\\
RF & Radio-frequency\\
RTD & Resistive thermal device\\
SC & Super conducting\\
SM & Standard model of particle physics\\
SRIM & Stopping and Range of Ions in Matter\\
$T_{1/2}$ & Half-life\\
TPC & Time projection chamber\\
TRIM & Transport of Ions in Matter\\

\end{tabular}
}

\begin{adjustwidth}{-\extralength}{0cm}
\printendnotes[custom] 

\reftitle{References}





\PublishersNote{}
\end{adjustwidth}
\end{document}